\documentclass[aps,prb,reprint,showpacs,superscriptaddress]{revtex4-1}
\usepackage[normalem]{ulem}
\usepackage[french]{babel}
\usepackage{relsize}
\usepackage[breaklinks=true,colorlinks,citecolor=blue,linkcolor=blue,urlcolor=blue]{hyperref}
\usepackage{graphicx,dcolumn,bm,multirow,hyperref,mathtools,braket,color,url,epstopdf,amssymb,amsmath}
\usepackage{xcolor}
\usepackage{graphicx}
\usepackage{float}                
\usepackage[section]{placeins} 
\usepackage[normalem]{ulem}
\usepackage{dcolumn}
\usepackage{bm}

\usepackage[normalem]{ulem}
\usepackage{soul}

\usepackage{multirow}
\usepackage[mathscr]{euscript}
\usepackage{gensymb}
\newcommand{\orcid}[1]{\href{https://orcid.org/#1}{\textcolor[HTML]{A6CE39}{\aiOrcid}}}

\DeclareSymbolFont{rsfs}{U}{rsfs}{m}{n}
\DeclareSymbolFontAlphabet{\mathscrsfs}{rsfs}

\preprint{APS/123-QED}

\begin{document}

\title{Direct visualization of gate-tunable flat bands in twisted double bilayer graphene}

\author{Souvik Sasmal}
\thanks{S. Sasmal and R. Muzzio contributed equally to this work.}
\affiliation{Department of Physics, Carnegie Mellon University, Pittsburgh, USA.}

\author{Ryan Muzzio}
\thanks{S. Sasmal and R. Muzzio contributed equally to this work.}
\affiliation{Department of Physics, Carnegie Mellon University, Pittsburgh, USA.}

\author{Ahmed Khalifa }
\affiliation{Department of Physics, Carnegie Mellon University, Pittsburgh, USA.}

\author{Paulina Majchrzak}
\affiliation{Department of Applied Physics, Stanford University, Stanford, CA 94305, USA}

\author{Alfred J.H. Jones}
\affiliation{Department of Physics and Astronomy, Aarhus University, 8000, Aarhus C, Denmark}
\author{I-Hsuan Kao}

\affiliation{Department of Physics, Carnegie Mellon University, Pittsburgh, USA.}

\author{Kenji Watanabe} 
\affiliation{Research Center for Electronic and Optical Materials, National Institute for Materials Science, 1-1 Namiki, Tsukuba 305-0044, Japan}

\author{Takashi Taniguchi}
\affiliation{Research Center for Materials Nanoarchitectonics, National Institute for Materials Science,  1-1 Namiki, Tsukuba 305-0044, Japan}

\author{ Simranjeet Singh }
\affiliation{Department of Physics, Carnegie Mellon University, Pittsburgh, USA.}

\author{Eli Rotenberg}
\affiliation{Advanced Light Source, E. O. Lawrence Berkeley National Laboratory, Berkeley, CA, 94720, USA}

\author{Aaron Bostwick}
\affiliation{Advanced Light Source, E. O. Lawrence Berkeley National Laboratory, Berkeley, CA, 94720, USA}

\author{Chris Jozwiak}
\affiliation{Advanced Light Source, E. O. Lawrence Berkeley National Laboratory, Berkeley, CA, 94720, USA}

\author{Søren Ulstrup}
\affiliation{Department of Physics and Astronomy, Aarhus University, 8000, Aarhus C, Denmark}

\author{Shubhayu Chatterjee}
\affiliation{Department of Physics, Carnegie Mellon University, Pittsburgh, USA.}

\author{ Jyoti Katoch}
\email{jkatoch@andrew.cmu.edu\\}
\affiliation{Department of Physics, Carnegie Mellon University, Pittsburgh, USA.}

\begin{abstract}
The symmetry-broken correlated states in twisted double bilayer graphene (TDBG) can be tuned via several external knobs, including twist angle, displacement field, and carrier density. However, a direct, momentum-resolved characterization of how these parameters reshape the flat-band structure remains limited.
In this study, we employ micro focused angle-resolved photoemission spectroscopy to investigate the flat-band dispersion of TDBG at a twist angle of 
$1.6^\circ$, systematically varying the displacement field and carrier density via electrostatic gating. We directly observe multiple flat moir\'e minibands near charge neutrality, including a flat remote valence band residing below the low-energy flat-band manifold. Furthermore, the dominant Coulomb repulsive energy over the flat-band bandwidth suggests favorable conditions for the emergence of interaction-driven correlated phenomena in TDBG. These findings establish that the formation and evolution of flat bands in TDBG arises from the interplay between the electron filling and the displacement field.

\end{abstract}

\maketitle


\begin{figure*}
\includegraphics[width=0.9\textwidth]{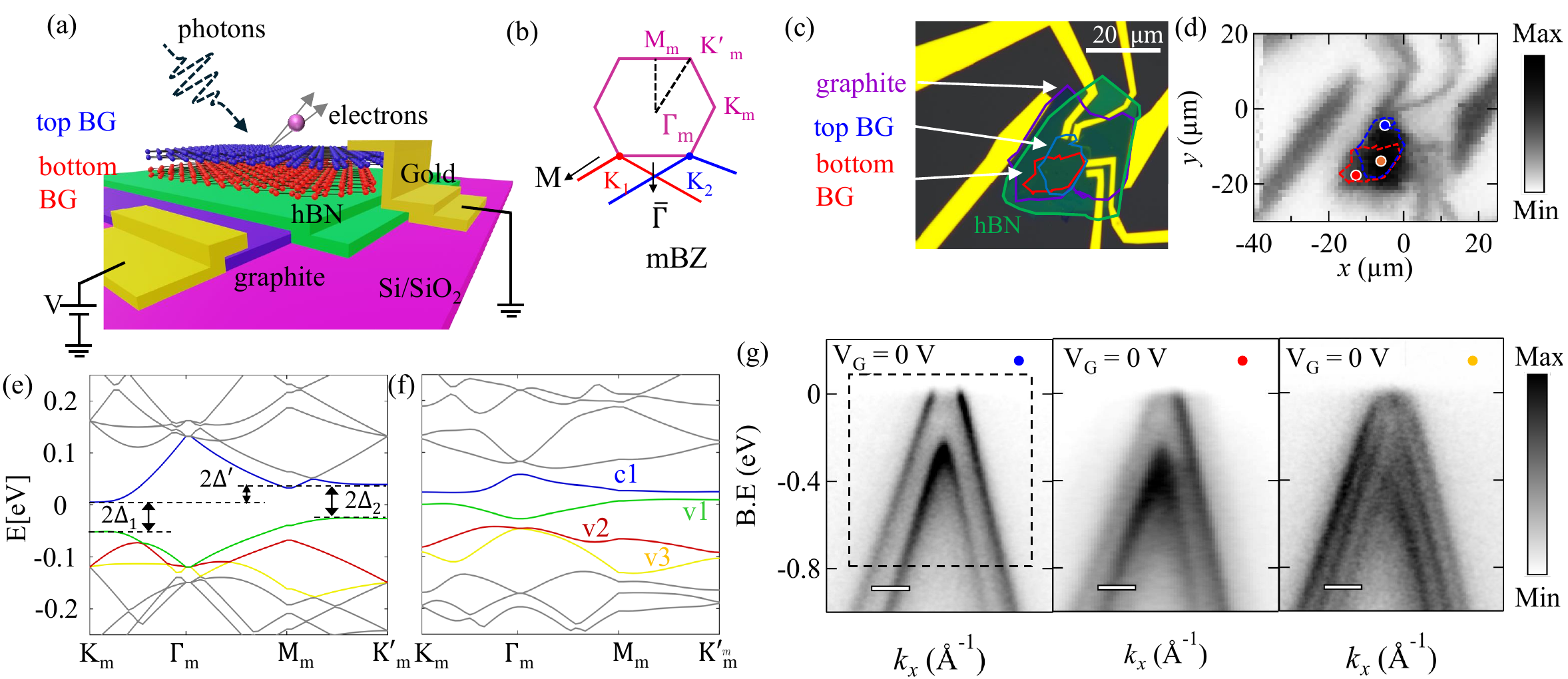}
\caption{(a) Schematic diagram of TDBG/hBN/graphite on SiO$_2$ substrate. It illustrates a photoemission setup with a microfocus beam and gate voltage is applied on graphite. (b) The purple hexagon geometry of the mini Brillouin zone (mBZ) originated from top BG (blue hexagon) and bottom BG (red hexagon). The blue and red dots mark the locations of originals ${\rm{K_2}}$-point from top BG and ${\rm{K_1}}$ point from bottom BG. mBZ high symmetry points $\Gamma_{\rm{m}}$, $\rm{K}_{\rm{m}}$, and $\rm{M}_{\rm{m}}$ are indicated in purple hexagon.(c) Optical microscope image of the device. Top BG (blue), bottom BG (red), hBN (green), and graphite (purple) regions are indicated by different color lines. (d) Spatially dependent ARPES intensity integrated over energy and momentum within the dashed box in the left panel of (g). This image is generated from real space mapping the valence band spectra and integrating the signal. (e) The theoretical band structure of TDBG in the absence of interlayer interaction. (f) The perturbative interband hybridization through the interlayer tunneling. (g) Energy-momentum dispersion of top BG, bottom BG, and TDBG at $\rm{V_G} = 0$~V. Scale bars = 0.1 \AA$^{-1}$} 
\label{xyscan}
\end{figure*}

\emph{Introduction.---}
Twisted double bilayer graphene (TDBG), composed of two Bernal-stacked bilayer graphene sheets with a relative twist angle, has emerged as a tunable platform for exploring flat band driven correlated physics without the stringent twist angle constraints of magic-angle twisted bilayer graphene (TBG)~\cite{Chen2019,PhysRevB.99.235417,liu2020tunable,Cao2020,FengWangTBG,Li2024,101126scienceaav1910,Park2021,101126sciadvaaw9770,PhysRevX.8.031089}. TDBG features flat bands at the Fermi level near a twist angle of $\theta \approx 1.30^\circ$ \cite{Shen2020,Lee2019}. An external electric field, induced via electrostatic gating, shifts the band edges of the two bilayer graphene (BG) sheets relative to each other, and induces band gaps between the narrow bands near charge neutrality.
Within the continuum non-interacting model of TDBG, the reduced kinetic energy of electronic states in the flat-bands (via suppresion of band-width $W$) should amplify the flat-band projected Coulomb interaction ($U/W \gtrsim 1$) and lead to intriguing correlated electron phases~\cite{Chen2019,PhysRevB.99.235417}. However, in reality, the filled electronic states and the applied displacement field will both renormalize the band-width, making it important to characterize these bands via direct imaging of their dispersion.

In this letter, we report the momentum-resolved tunability of moiré induced correlated states in TDBG utilizing non-invasive in operando angle-resolved photoemission spectroscopy with micrometer spatial resolution (microARPES) {~\cite{Katoch2018,RyanPRB,PhysRevMaterials.2.074006,AlfredAdvaceMaterial}. 
By characterizing the flat moiré minibands as a function of carrier density and displacement field, both tunable via applied gate voltages, we establish robustness of these flat bands across the moiré Brillouin zone. Interestingly, on increasing the displacement field, we observe not only the predicted emergence of flattened bottom conduction and top valence band (reduction of bandwidth) near $E_{\rm{F}}$, but the valence band at higher binding energy also appears to be flattened, opening up the possibility of complex correlated states in the so-called remote bands in this system. 
By directly visualizing the filling-dependent bandwidth, we establish that the Coulomb interaction energy scale remains dominant over the kinetic energy and therefore plays a key role in determining the electronic ground state. Our results thus provide crucial information for determining the correlated electronic landscape of TDBG. Further, our lack of observation of replica flat bands with uniform spacing, in sharp contrast to ARPES and STM studies of superconducting TBG with strong electron-boson couping~\cite{Chen2024}, also indicates that the electron-phonon coupling in TDBG is weaker, enhancing our understanding of the absence of definitive signatures of superconductivity in TDBG~\cite{liu2020tunable,Cao2020}.

\begin{figure*}[]
\centering
\begin{minipage}[c]{0.75\textwidth}
\includegraphics[width=1\textwidth]{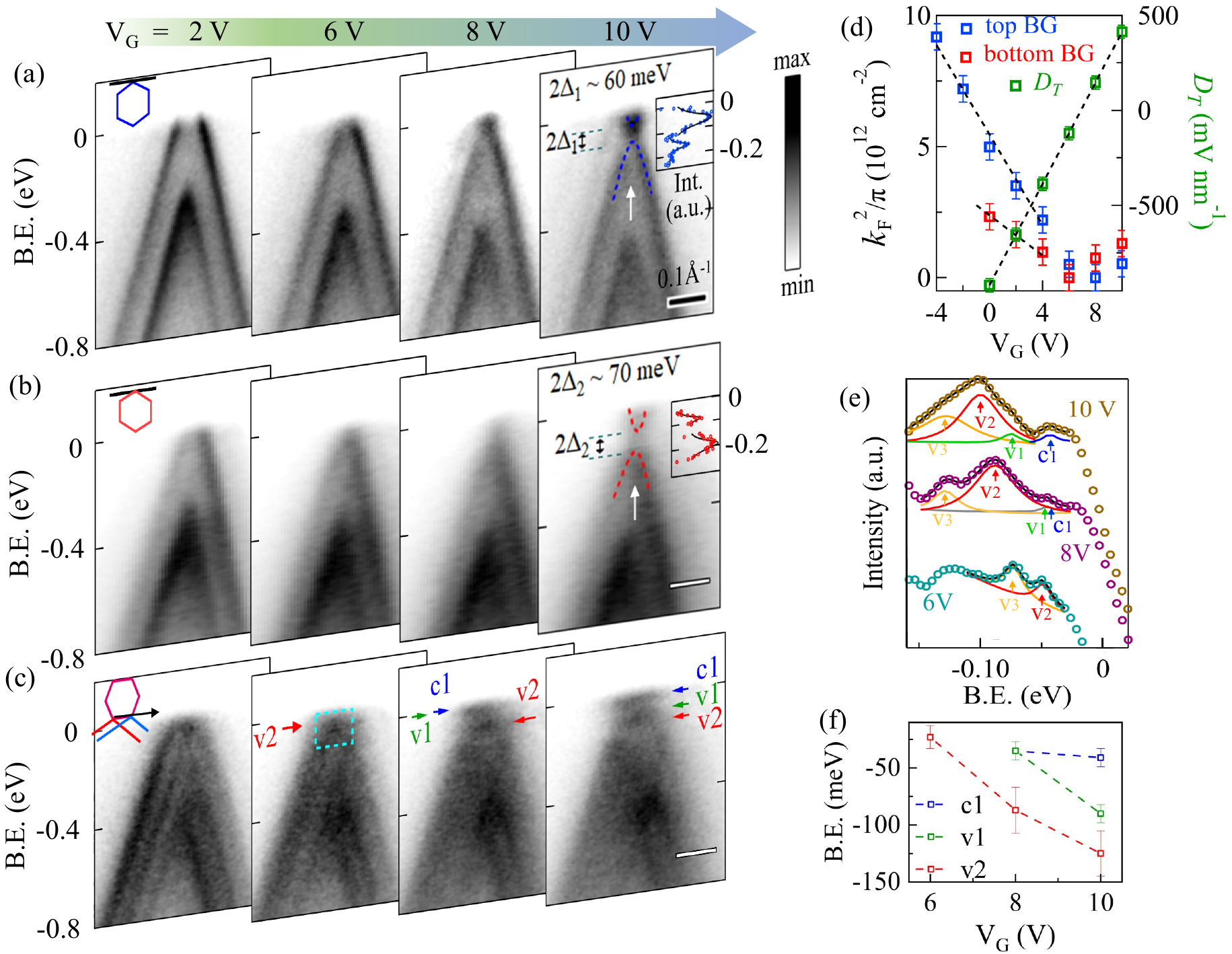}
\end{minipage}\hfill
\begin{minipage}[c]{0.25\textwidth}
\caption{Snapshot of energy-momentum dispersion of (a) top BG (top panel), (b) bottom BG (middle panel), and (c) TDBG (bottom panel) at different $V_G=$ 2 V, 6 V, 8 V, and 10 V (from left to right). The scale bar indicates 0.1~\AA$^{-1}$. Scale bars = 0.1 \AA$^{-1}$. Insets show EDC along arrow. (d) Carrier densities, for both top and bottom BGs, are determined as $k_{\rm{F}}^2/\pi$. $k_{\rm{F}}$ is the Fermi momentum vector. The calculated displacement field (${D}_T$), determined from the carrier densities, is shown on the right axis. (e) The average EDCs extracted over the flat band region for different $V_G$. The solid black line indicates Lorentzian peaks fit at the $V_G = 6, 8, \rm{and}~10~\rm{V}$. Blue, green, red, and yellow arrows and lines demarcate peak positions and fittings corresponding to c1, v1, v2, and v3, respectively. (f) The fitted peak positions of c1, v1, and v2 from the EDC analysis at different $V_G$.}
\label{gatetuned}
\end{minipage}
\end{figure*}


\emph{Experiment.---}
Fig. ~\ref{xyscan}(a) illustrates the overall schematic of the ARPES measurements on the TDBG device, which were performed using x-ray capillary ($Sigray~Inc.$) with a spatial resolution of (1.8 ± 0.3)$\mu$m. The TDBG heterostructure was fabricated using the ``tear and stack" method as described in the supplemental Materials (SM) Fig.~S1 in Sec.~A~\cite{supply}. Fig. ~\ref{xyscan}(b) shows the mBZ (purple hexagon) created by the top BG (blue) and bottom BG (red), with high-symmetry points marked within the mBZ. The studied TDBG device consists of two twisted BG sheets supported on hexagonal boron nitride (hBN) flake placed over a graphite gate, with gold electrodes patterned on a SiO$_2$/Si substrate as shown in the optical microscope image in Fig. ~\ref{xyscan}(c). All dispersions obtained on individual bilayer graphene and TDBG were measured at $\sim$ 20~K, with a photon energy of 90~eV and pass energy 30~eV. Next, we obtain a four-dimensional (4D) data set containing the (E, $k$, $x$, $y$)- dependent spatial photoemission intensity map over the stacking area as captured in the optical image. The (E,$k$)- integrated intensity within the spectral region around the K-point in Fig. ~\ref{xyscan} (g) is used to generate the photoemission intensity map shown in Fig. ~\ref{xyscan}(d). From momentum-resolved energy slices at $E_{\rm{F}}$, the twist angle was determined to be approximately $1.6^{\circ} \pm 0.2^{\circ}$ between the two bilayer graphene sheets (details in SM Fig.~S3 in Ref.~\cite{supply}). Fig.~\ref{xyscan}(e) and (f) represent the calculated electronic structure of 1.6$\degree$ twist TDBG in the absence and presence of interlayer hybridization, respectively. In Fig.~\ref{xyscan}(e), we show the band gaps $2\Delta_1$ and $2\Delta_2$ in the uncoupled bands of the individual BG sheets, and the band offset $2\Delta'$ between these, due to the perpendicular electric field.  
On the other hand, Fig.~\ref{xyscan}(f) shows the emergence of flat bands from perturbative interband hybridization, caused by interlayer tunneling where the two BG sheets overlap.

\emph{Results.---} In Fig.~\ref{xyscan}(g), we observe two $\pi$ bands in the energy-momentum dispersion around K-points for both the top and bottom BG (measured at blue and red colored dot region of the spatial intensity map~Fig.~\ref{xyscan}(d)). In contrast, the twisted region exhibited hybridized bands at the Fermi level ($E_{\rm{F}}$) (measured in the yellow-colored dot region of the spatial intensity map) as shown in Fig. ~\ref{xyscan}(g). We observe that both the top and bottom BGs are $p$- doped with a different degree of doping at gate voltage ($V_{\rm{G}}$) of 0 V (discussed later in detail). This unintentional $p$-doping of graphene and bilayer graphene has been previously observed in transport and ARPES studies and can be attributed to adsorbents such as organic residue, water, oxygen, and weak metal contact~\cite{PhysRevMaterials.3.110301,10.1063/5.0164903,10.1063/5.0161160,PhysRevLett.101.026803}.


The selected series of snapshots of E(k) dispersions at different $\rm{V_G}$ are analyzed further in detail for top BG (Blue spot in Fig.~\ref{xyscan}(b)), bottom BG (red spot in Fig.~\ref{xyscan}(b)) and TDBG (yellow spot in Fig.~\ref{xyscan}(b)) as shown in Fig.~\ref{gatetuned}(a),(b) and (c), respectively. 
For each of these panels, we demonstrate the accessibility of the valence and conduction bands by tuning the charge carriers from $p$- to $n$-type by applying ${\rm{V_{G}}}$ in the range from 2 V to 10 V. For bottom BG (top BG), the conduction band appears at $V_{G} \geq 6 \rm{V}$ ($V_{G} \geq 8 \rm{V}$). 
Applying a gate voltage not only tunes the carrier density, but also applies a displacement field to the system. This field leads to band gaps and band offset within each BG layer and also between top and bottom BG layers. To estimate the displacement field induced band-gaps of the non-interacting TDBG, we extract the displacement field induced band-gaps by fitting a parabolic dispersion for BGs with dashed lines at 10~V in Fig.~\ref{gatetuned}(a) and (b). At 10~V, the external field creates a maximum gap of $2\Delta_1 ~\sim~60$~meV for top BG and $2\Delta_2 ~\sim70$~meV for bottom BG. The obtained band gap uncertainty 25 meV is estimated from the linewidths of the Lorentzian fits. Here, the band gap defines the interlayer potential difference in each BG. The band offset defines the potential difference between the two BG sheets and it is determined as a shift of the conduction band of the bottom BG with respect to the top BG. Here, the band offset $2\Delta'~\sim40~\rm{meV}$. This variation in band offset is more likely due to inhomogeneous doping and local variations in the displacement field across the device, which in turn influence the band hybridization observed in the TDBG, see Fig.~\ref{gatetuned}(c).

Next, we investigate the tunability of disperions in the three regions (top BG, bottom BG, and TDBG) of interest with the carrier concentration and displacement field using $\rm{V_G}$ in our device. For bilayer graphene, an approximate low-energy electronic dispersion~\cite{McCann_2011,raza2012graphene,WilsonTG} in the density range of interest is given by 
\begin{equation}
    E_{\rm{B.E}} = E_D + \frac{g_1}{2}\left(\sqrt{1+\frac{4v^2}{g_1^2}k^2}-1 \right)
\label{equ1}
\end{equation}
Here, $g_1$ is the interlayer tunneling, $v$ is the band velocity and $E_D$ is the charge neutrality point.
 
This formula interpolates between the linear dispersion (large momenta at $E_{\rm{F}}$, $V_G\leq4$~V) to quadratic dispersion (small momenta at $E_{\rm{F}}$, $V_G > 6~\rm{V}$). For $V_G\leq4$~V, the carrier densities (n), for both top and bottom BGs, are given as $n = k_{\rm{F}}^2/\pi$, where $k_{\rm{F}}$ is the Fermi momentum vector. On the other hand, for $V_G>6~\rm{V}$, the momenta at $E_{\rm{F}}$ is small, leading to a low carrier density. In this region, $v k/g_1 \ll 1$, and the dispersion is approximately quadratic. In this limit, the bilayer carrier density takes a simple form: $n \approx \frac{g_1 E_D}{\pi v^2}$~\cite{WilsonTG}. As shown in Fig.~\ref{gatetuned}(d), $k_F^2/\pi$ follows linear behavior for $V_G\leq4$~V and begins to deviate above 4~V.

\begin{figure*}[t!]
\centering
\begin{minipage}[c]{0.68\textwidth}
\includegraphics[width=1\textwidth]{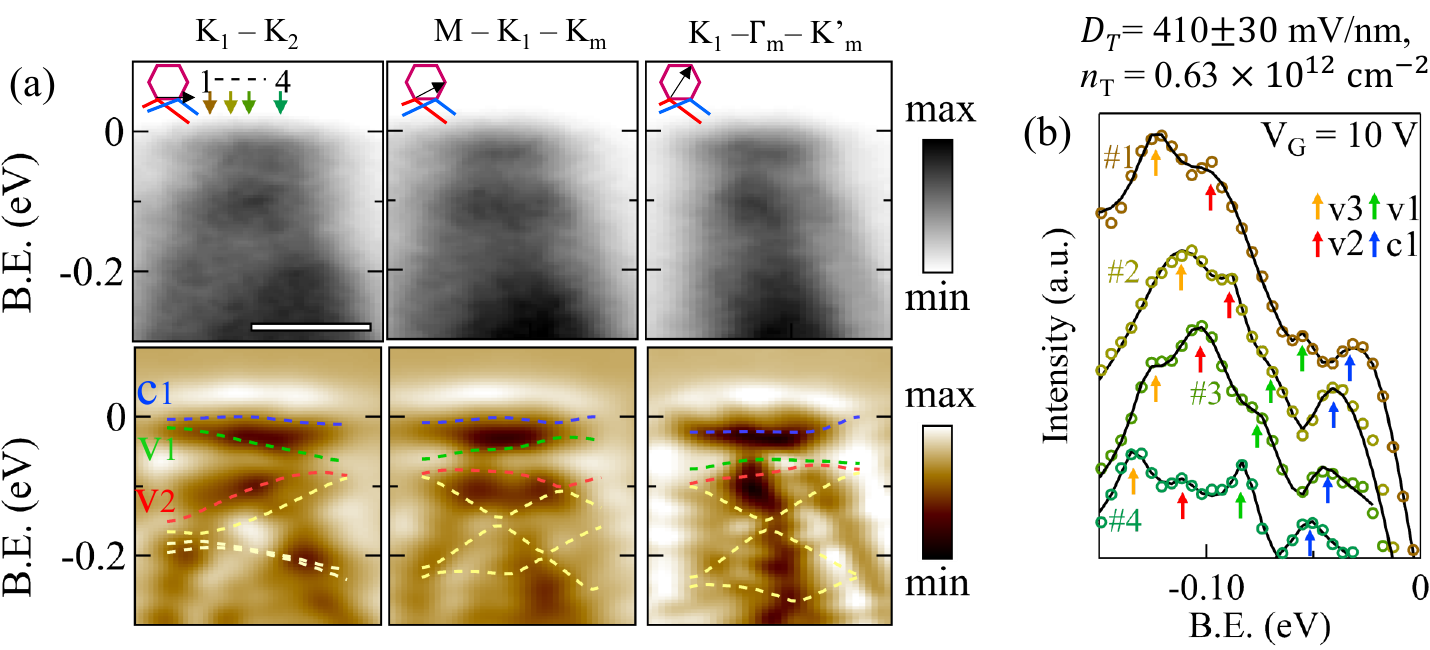}
\end{minipage}\hfill
\begin{minipage}[c]{0.3\textwidth}
\caption{(a) The top panel presents the energy-momentum dispersion of TDBG under an applied gate voltage of 10 V along different high-symmetry directions in the moiré Brillouin zone (mBZ). The cut directions are indicated by arrows in the purple hexagon (mBZ schematic) at the top-left corner. Scale bar = 0.1~\AA$^{-1}$. The bottom panel shows the corresponding second-order derivative plots. The calculated band positions—c1, v1, v2, and v3—are represented by blue, green, red, and yellow dashed lines, respectively. (b) EDCs $\#1 – \#4$, extracted along the $\rm{K_1}$ – $\rm{K_2}$ cut direction from the top-left figure, illustrate the band positions. The corresponding color-coded arrows indicate the alignment of the extracted bands.}
\label{directional_Ekx}
\end{minipage}
\end{figure*}

 The peak positions of Lorentzian fit of momentum dispersion curves (MDCs) at $E_{\rm{F}}$ correspond to the Fermi wave vectors of the electronic states, and the Fermi momentum vector ($k_{\rm{F}}$) is determined as half of the distance between the peaks. 
SM Fig.~S4 in Ref.~\cite{supply} shows the MDCs cuts at different gate voltages. The extracted gate voltage-dependent carrier density ($n$), for both top- and bottom- BG, is shown in Fig.~\ref{gatetuned}(d). The p-type carrier density varies linearly with the $\mathrm{V_G}$ and is given by: $n = \frac{C}{q} (\mathrm{V_G - V_0})$, for $\mathrm{V_G} \leq 4V$. Here, $C$ represents the estimated capacitance per unit area, and $\mathrm{V_0}$ is a constant shift. The linear fits of this equation for both the top and bottom BG sheets are shown as black dotted lines in Fig.~\ref{gatetuned}(d). From the fit, for top BG (bottom BG), the extracted capacitance value $C_{\text{top}} = 120 \pm 7 \text{ nF}/\text{cm}^2 \quad (C_{\text{bottom}} = 68 \pm 5 \text{ nF}/\text{cm}^2),$  while the corresponding shift voltage $V_{0,\text{top}} = 7.3 \pm 0.1~\mathrm{V} \quad (V_{0,\text{bottom}} = 6.1 \pm 0.1$~V). It reveals that at ${\mathrm{V_G}}=$ 0~V, the top BG (bottom BG) sheets are p-doped with charge- carrier densities of $5.4 \pm 0.1 \times 10^{12}~\mathrm{cm}^{-2}$ ($2.3 \pm 0.1 \times 10^{12}~\mathrm{cm}^{-2}$), resulting in an unintentional displacement field (${D}_T$) in the TDBG region. At $\mathrm{V_G} = 10~V$, both the conduction and valence bands of top- and bottom- BGs are fitted with Eq.~\ref{equ1}, shown in right panel in Fig.~\ref{gatetuned}(a,b). The estimated $g_1 = 0.3~\mathrm{eV}$ ($0.3~\mathrm{eV}$), $4v^2 = 484.4~\mathrm{eV^2 \AA^2}$ ($310.4~\mathrm{eV^2 \AA^2}$) and $E_D = -0.09~\mathrm{eV}~(-0.05~eV)$ for bottom- (top-) BG conduction band.   So, at $\mathrm{V_G} = 10~V$, the calculated carrier density for bottom-BG (top-BG) is $\sim 0.71\times10^{12}~\mathrm{cm^{-2}}$ ($\sim 0.57\times10^{12}~\mathrm{cm^{-2}}$). From our experimental data we cannot extract the charge carrier density in the TDBG region but we can estimate it by assuming it to be average value of the charge density calculated in the top and bottom BG layer. The estimated average carrier density $ n_{\mathrm{ave}} =0.63\times10^{12}~\mathrm{cm^{-2}}$ at $\mathrm{V_G} = 10~V$. Assuming that the charge neutrality of TDBG occurs at $V_{CN} = (V_{0,\rm{top}}+V_{0,\rm{bottom}})/2  \approx~6.9\pm0.1~\rm{V}$, relative permittivity of hBN $\epsilon_{BN}\sim 3 - 5$ and thickness $d \sim$ 15 nm (confirmed via atomic force microscopy, Fig. S2 in Ref.~\cite{supply}), the estimated displacement field is $D_T \sim$ $\epsilon_{BN}(\rm{V_G} - V_{CN})/2d~V/nm.$ A detailed discussion of relative orientation and the dielectric environment of hBN are shown in SM Fig.~S5 in Ref.~\cite{supply}. The right axis in  Fig.~\ref{gatetuned} (d) shows $D_T$ as a function of $V_G$ and a linear fit, drawn by a dashed line, indicates the linear behavior of $D_T$ with $V_{\rm{G}}$.

We now focus on the effects of carrier density and displacement field variation on the moir\'e bands of TDBG. The average energy distribution curves (EDCs) cuts for TDBG, extracted over the flat band region, at different $V_{\rm{G}}$ are shown in Fig.~\ref{gatetuned}(e). We observe that at $V_G = 6~\rm{V}$ ( carrier density $n_{\rm{ave}}\sim - 0.25\times 10^{12}~\rm{cm}^{-2}$ and ${D}_T \sim -120\pm30~\rm{mV~nm^{-1}}$) the valence bands v2 and v3 exhibit peak positions—corresponding to band maxima—as shown by the red and yellow arrows, respectively, in the fitted curves in Fig.~\ref{gatetuned}(e). At $V_G = 8~\rm{V}$ ($n_{\rm{ave}} = 0.5 \times 10^{12}~\rm{cm}^{-2}$, ${D}_T \sim 150\pm30~\rm{mV~nm^{-1}}$), the band dispersion—indicated by peak positions—shows that the primary conduction (c1, blue arrow) and valence (v1, green arrow) bands appear merged into a single flat band near B.E. $-45~\rm{meV}$, as seen in Fig.~\ref{gatetuned}(e). A second valence band v2 (red arrow) is observed at B.E. $-87~\rm{meV}$.
The band structure, especially c1 and v1 bands, evolve as we tune ${D}_T$ and $n_{\rm{ave}}$ by increasing $V_G$ (see Fig.~\ref{gatetuned}(e)). At maximum measured displacment field, ${D}_T~\sim410\pm30~\rm{mV~nm^{-1}}$ (and the $n_{\rm{ave}}$ is $\sim 0.63~\times~10^{12}~\rm{cm}^{-2}$) by applying $V_G = 10~\rm{V}$, we observe two distinct flat bands$-$conduction band c1 and valence band v1 at B.E. of $-42\pm13$~meV and $-74\pm12$~meV respectively, with a energy separation of $\sim30$~meV, in addition to valence band v2 at B.E. of $-98\pm20$~meV. These experimental results are consistent with previous transport studies~\cite{He2021,Su2023} and the recent scanning tunneling microscope (STM) experiment ~\cite{Liu2021} that show c1 and v1 appear to be flat bands and the gap between c1 and v1 increases with the increasing gate voltage. In transport measurements He $et~al$., reported the resistivity as a function of carrier density and displacement field phase diagram in TDBG which shows semimetal-to-semiconductor transition at the charge neutrality point with increasing $|{D}_T|$, as well as bandgaps at full filling of the valence and conduction bands~\cite{He2021}.  In these measurements an insulating state appears for displacement field $ \geq 250~\rm{mV~nm^{-1}}$ and carrier density $\approx 1.8 - 2.5 \times 10^{12}~\rm{cm}^{-2}$~~\cite{He2021,Su2023}. Similarly, in our case, we  observed flat like dispersion of valence and conduction bands v1 and c1 respectively with carrier density $0.63\times10^{12}~\rm{cm^{-2}}$ and $|{D}_T|~\sim~410\pm30~\rm{mV~nm^{-1}}$.

Additionally, to have a global view of the band structure, we take different energy-momentum cuts and analyze the band dispersions across the mBZ. The flat band feature should exist across the whole mBZ. Band dispersion along high symmetry directional cuts for TDBG at $V_G = 10~\rm{V}$ confirm flat band features are present over the whole mBZ  as seen in the top panel of Fig.~\ref{directional_Ekx} (a). 
The mBZ is indicated by a purple hexagon and the arrow indicates the energy-momentum dispersion cut direction in the mBZ. 
The corresponding second partial derivative plots are shown in the bottom panel of Fig.~\ref{directional_Ekx} (a). We compare our experimental results of gate-tuned band dispersion to the band structure obtained from the continuum model for TDBG~\cite{Haddadi2020,Lee2019}. The theoretically calculated bands c1 (blue), v1 (green), v2 (red), and v3 (yellow) are overlaid on the second partial derivative plots and indicated as dash lines (see Fig.~\ref{directional_Ekx}(a)). Our observation shows reasonable agreement with the continuum model away from the $\Gamma_m$ point (Fig.~\ref{directional_Ekx}(a), left and middle panels), while revealing significant interaction-induced renormalization near $\Gamma_m$ (Fig.~\ref{directional_Ekx}(a), right panel). A small discrepancy between theory and experiment may arise from uncertainties in the twist angle, applied electric field, dielectric screening, or related device-specific parameters. Fig.~\ref{directional_Ekx}(b) shows the EDCs at different momentum vectors indicated by arrows in Fig.~\ref{directional_Ekx}(a). Due to the high quality of the dispersions, it is possible to extract the bandwidths from our experimental data with back gate voltage. At $V_G = 10~\rm{V}$, the bandwidths ($W\rm{s}$) of c1,v1 and v2 are $\sim28\pm13$~meV, $\sim45\pm12$~meV, and $\sim30\pm20$~meV, respectively: these are reasonably consistent with the continuum model estimates of $39$~meV, $33$~meV and $54$~meV respectively. The present experimental arrangement partly limits the relatively broad line widths. These range of $W$s, with a BG interlayer potential difference $\sim 60~ \rm{meV}$ and band offset $\sim40\pm20~\rm{meV}$, is achieved by displacement field $\sim~410\pm30~\rm{mV~nm^{-1}}$ and carrier concentration $0.63\times10^{12}~\rm{cm^{-2}}$. Typically, a relatively narrow bandwidth on the order of $\sim$ 25 meV or smaller is achievable for moderate electric fields that introduce interlayer potential differences typically of a few tens of meV~\cite{PhysRevB.99.235417}. Here, the flat bands are dispersed over $\sim$~0.09~\AA$^{-1}$ which is about the width of $\rm{K_1 - \Gamma - K'_m}$ ($\lambda _m =\frac{8\pi}{\sqrt{3}a}~\rm{sin\left(\frac{\theta}{2}\right)}\approx$~0.083~\AA$^{-1}$) for TDBG (see Fig.~\ref{directional_Ekx}(a)). This indicates the observed flat band is approximately the same size of mBZ of TDBG. For the magic angle twisted bilayer graphene, the observed flat band (spread over $\sim0.1~\rm{\AA}^{-1}$ along $k_x$) is approximately twice the mBZ $\sim0.06~\rm{\AA}^{-1}$~\cite{Utama2021}.

In our case, considering the moire wave-length $4\pi/\sqrt3~\lambda_m~\sim 8~\rm{nm}$ and hBN dielectric $\epsilon_{BN}\sim 3 - 5$, the estimated on-site Coulomb repulsion energy 
$U~(=\frac{e^2}{4\pi\epsilon}~\frac{\lambda_m \sqrt{3}}{4 \pi})$
$\sim (40 ~- ~66) ~\rm{meV}$, which suggests $U/W \gtrsim 1$ in the device. 
Additionally, since the gate thickness of $d \approx 15$ nm is substantially larger than the moir\'e wavelength, we do not expect $U$ to be significantly affected by gate-screening.
We note that a similar value of $U$ has been estimated in 
a transport study on TDBG device~\cite{PhysRevB.101.125428}. 
This dominant on-site Coulomb repulsion may raise the possibility of complex correlated states in the flat bands. 
Similarly, a larger Coulomb interaction compared to bandwidth has been established in transport measurements in gate-tuned multilayer twisted graphene systems, indicating a possible correlated insulating state~\cite{Chen2019,PhysRevB.101.125428,101126scienceabg3036}. 



In our case, at $\rm{V_G = 10~V}$, we observe that the bandwidth of c1 is smaller than that of v1, favoring isolation of the conduction band and indicating broken particle-hole symmetry, consistent with expectations from a non-zero displacement field. Here, a strong applied displacement field, generated due to a single-gated geometry, may lead to breaking particle--hole symmetry in the system and tend to favor the isolation of conduction flat bands over valence bands 
This smaller bandwidth of the flat bands suggests reduced screening and stronger Coulomb interactions. 
Interestingly, we also found that valence band v2 (at higher binding energy) has a narrower bandwidth compared to v1, suggesting flat remote valence band. Since the band gap between v1 and v2 is comparable to the interaction strength $U$ as estimated previously, our results also open the door to realizing multi-band correlated phases in TDBG. However, it does not predict whether the insulators at high filling correspond to remote band v2.
The Lorentzian fits of the intensity plots at $V_G = 8~\rm{V}$ are shown in SM Fig.~S6 in Ref.~\cite{supply}. 
The intensity distribution curves confirm the flat band contains overlapping c1 and v1 bands with bandwidth $\sim 40\pm15$~meV.


\emph{Summary.---}
Recently, momentum-resolved studies have revealed flat miniband dispersions in TDBG, shedding light on many-body effects in gated 2D heterostructures under varying carrier densities and displacement fields~\cite{WilsonTG, Jiang2025}. Despite these advances, important questions remain—particularly concerning the interplay between electron filling, displacement field, and the emergence of correlated states. In this work, we present a detailed momentum-resolved electronic structure of top BG, bottom BG and TDBG that consitute the overlapped region between top and bottom BG,  as a function of applied back-gate voltage. By separately probing the top and bottom BG components, we establish the evolution of charge carrier density and displacement field at different gate voltages, which gives an estimate of the displacement field and charge carrier density in the TDBG region. At $V_G = 8~\rm{V}$, we observe the emergence of a flat band near the Fermi level in TDBG, formed by overlapping conduction (c1) and valence (v1) states. Further increasing the gate voltage to $V_G = 10~\rm{V}$ adds electrons and enhances the displacement field, which lifts the degeneracy and increases the band gap between the low-energy conduction and valence bands. In addition to the low-energy flat-band manifold, we also observe a flat remote valence band at higher binding energy whose appearance correlates with increased displacement field. These tunable flat bands highlight the complex band evolution in TDBG and provide a versatile platform to explore interaction-driven phenomena and their interplay with non-trivial band topology, e.g., Chern bands in TDBG~\cite{Shen2020, PhysRevLett.133.246401, qc2g-qjg2, PhysRevB.99.235406, Perea-Causin2025} and other engineered 2D systems. Finally, we do not observe features resembling replica flat bands in TDBG at $V_G = 8~\rm{V}$ and $V_G = 10~\rm{V}$, in contrast to hBN-unaligned magic-angle TBG~\cite{Chen2024}, where such features are associated with strong electron-phonon coupling and were observed only in TBG samples exhibting superconductivity. The abscence of strong electron-phonon coupling in our TDBG combined with lack of clear signatures of superconductivity in transport measurements ~\cite{liu2020tunable, Cao2020}, hint towards a conventional phonon-induced mechanism for superconductivity in moir\'e graphene. Thus, our work motivates further investigation of the interplay of flat-band electrons and phonons to understand the origin of superconductivity in these remarkable materials. 

\emph{Acknowledgements--} This work was mainly supported by the U.S. Department Office of Science, Office of Basic Sciences, of the U.S. Department of Energy under Award No. DE-SC0020323  and DE-SC0025490. Partial support was provided by NSF Early Career under award number Award No. 2339309 (device fabrication and characterization). S.S. acknowledges the support from National Science Foundation under grant DMR-2210510. K.W. and T.T. acknowledge support from the JSPS KAKENHI (Grant Numbers 21H05233 and 23H02052) , the CREST (JPMJCR24A5), JST and World Premier International Research Center Initiative (WPI), MEXT, Japan. S.U. acknowledges to the Novo Nordisk Foundation (Project Grant NNF22OC0079960). This research used resources of the Advanced Light Source, which is a DOE Office of Science User Facility under contract no. DE-AC02-05CH11231.

\emph{Data availability--} The data are available upon reasonable request from the authors.


\begin{thebibliography}{42}%
\makeatletter
\providecommand \@ifxundefined [1]{%
 \@ifx{#1\undefined}
}%
\providecommand \@ifnum [1]{%
 \ifnum #1\expandafter \@firstoftwo
 \else \expandafter \@secondoftwo
 \fi
}%
\providecommand \@ifx [1]{%
 \ifx #1\expandafter \@firstoftwo
 \else \expandafter \@secondoftwo
 \fi
}%
\providecommand \natexlab [1]{#1}%
\providecommand \enquote  [1]{``#1''}%
\providecommand \bibnamefont  [1]{#1}%
\providecommand \bibfnamefont [1]{#1}%
\providecommand \citenamefont [1]{#1}%
\providecommand \href@noop [0]{\@secondoftwo}%
\providecommand \href [0]{\begingroup \@sanitize@url \@href}%
\providecommand \@href[1]{\@@startlink{#1}\@@href}%
\providecommand \@@href[1]{\endgroup#1\@@endlink}%
\providecommand \@sanitize@url [0]{\catcode `\\12\catcode `\$12\catcode `\&12\catcode `\#12\catcode `\^12\catcode `\_12\catcode `\%12\relax}%
\providecommand \@@startlink[1]{}%
\providecommand \@@endlink[0]{}%
\providecommand \url  [0]{\begingroup\@sanitize@url \@url }%
\providecommand \@url [1]{\endgroup\@href {#1}{\urlprefix }}%
\providecommand \urlprefix  [0]{URL }%
\providecommand \Eprint [0]{\href }%
\providecommand \doibase [0]{http://dx.doi.org/}%
\providecommand \selectlanguage [0]{\@gobble}%
\providecommand \bibinfo  [0]{\@secondoftwo}%
\providecommand \bibfield  [0]{\@secondoftwo}%
\providecommand \translation [1]{[#1]}%
\providecommand \BibitemOpen [0]{}%
\providecommand \bibitemStop [0]{}%
\providecommand \bibitemNoStop [0]{.\EOS\space}%
\providecommand \EOS [0]{\spacefactor3000\relax}%
\providecommand \BibitemShut  [1]{\csname bibitem#1\endcsname}%
\let\auto@bib@innerbib\@empty
\bibitem [{\citenamefont {Chen}\ \emph {et~al.}(2019)\citenamefont {Chen}, \citenamefont {Jiang}, \citenamefont {Wu}, \citenamefont {Lyu}, \citenamefont {Li}, \citenamefont {Chittari}, \citenamefont {Watanabe}, \citenamefont {Taniguchi}, \citenamefont {Shi}, \citenamefont {Jung}, \citenamefont {Zhang},\ and\ \citenamefont {Wang}}]{Chen2019}%
  \BibitemOpen
  \bibfield  {author} {\bibinfo {author} {\bibfnamefont {G.}~\bibnamefont {Chen}}, \bibinfo {author} {\bibfnamefont {L.}~\bibnamefont {Jiang}}, \bibinfo {author} {\bibfnamefont {S.}~\bibnamefont {Wu}}, \bibinfo {author} {\bibfnamefont {B.}~\bibnamefont {Lyu}}, \bibinfo {author} {\bibfnamefont {H.}~\bibnamefont {Li}}, \bibinfo {author} {\bibfnamefont {B.~L.}\ \bibnamefont {Chittari}}, \bibinfo {author} {\bibfnamefont {K.}~\bibnamefont {Watanabe}}, \bibinfo {author} {\bibfnamefont {T.}~\bibnamefont {Taniguchi}}, \bibinfo {author} {\bibfnamefont {Z.}~\bibnamefont {Shi}}, \bibinfo {author} {\bibfnamefont {J.}~\bibnamefont {Jung}}, \bibinfo {author} {\bibfnamefont {Y.}~\bibnamefont {Zhang}}, \ and\ \bibinfo {author} {\bibfnamefont {F.}~\bibnamefont {Wang}},\ }\href {\doibase 10.1038/s41567-018-0387-2} {\bibfield  {journal} {\bibinfo  {journal} {Nat. Phys.}\ }\textbf {\bibinfo {volume} {15}},\ \bibinfo {pages} {237} (\bibinfo {year} {2019})}\BibitemShut {NoStop}%
\bibitem [{\citenamefont {Chebrolu}\ \emph {et~al.}(2019)\citenamefont {Chebrolu}, \citenamefont {Chittari},\ and\ \citenamefont {Jung}}]{PhysRevB.99.235417}%
  \BibitemOpen
  \bibfield  {author} {\bibinfo {author} {\bibfnamefont {N.~R.}\ \bibnamefont {Chebrolu}}, \bibinfo {author} {\bibfnamefont {B.~L.}\ \bibnamefont {Chittari}}, \ and\ \bibinfo {author} {\bibfnamefont {J.}~\bibnamefont {Jung}},\ }\href {\doibase 10.1103/PhysRevB.99.235417} {\bibfield  {journal} {\bibinfo  {journal} {Phys. Rev. B}\ }\textbf {\bibinfo {volume} {99}},\ \bibinfo {pages} {235417} (\bibinfo {year} {2019})}\BibitemShut {NoStop}%
\bibitem [{\citenamefont {Liu}\ \emph {et~al.}(2020)\citenamefont {Liu}, \citenamefont {Hao}, \citenamefont {Khalaf}, \citenamefont {Lee}, \citenamefont {Ronen}, \citenamefont {Yoo}, \citenamefont {Haei~Najafabadi}, \citenamefont {Watanabe}, \citenamefont {Taniguchi}, \citenamefont {Vishwanath} \emph {et~al.}}]{liu2020tunable}%
  \BibitemOpen
  \bibfield  {author} {\bibinfo {author} {\bibfnamefont {X.}~\bibnamefont {Liu}}, \bibinfo {author} {\bibfnamefont {Z.}~\bibnamefont {Hao}}, \bibinfo {author} {\bibfnamefont {E.}~\bibnamefont {Khalaf}}, \bibinfo {author} {\bibfnamefont {J.~Y.}\ \bibnamefont {Lee}}, \bibinfo {author} {\bibfnamefont {Y.}~\bibnamefont {Ronen}}, \bibinfo {author} {\bibfnamefont {H.}~\bibnamefont {Yoo}}, \bibinfo {author} {\bibfnamefont {D.}~\bibnamefont {Haei~Najafabadi}}, \bibinfo {author} {\bibfnamefont {K.}~\bibnamefont {Watanabe}}, \bibinfo {author} {\bibfnamefont {T.}~\bibnamefont {Taniguchi}}, \bibinfo {author} {\bibfnamefont {A.}~\bibnamefont {Vishwanath}},  \emph {et~al.},\ }\href {\doibase 10.1038/s41586-020-2458-7} {\bibfield  {journal} {\bibinfo  {journal} {Nature}\ }\textbf {\bibinfo {volume} {583}},\ \bibinfo {pages} {221} (\bibinfo {year} {2020})}\BibitemShut {NoStop}%
\bibitem [{\citenamefont {Cao}\ \emph {et~al.}(2020)\citenamefont {Cao}, \citenamefont {Rodan-Legrain}, \citenamefont {Rubies-Bigorda}, \citenamefont {Park}, \citenamefont {Watanabe}, \citenamefont {Taniguchi},\ and\ \citenamefont {Jarillo-Herrero}}]{Cao2020}%
  \BibitemOpen
  \bibfield  {author} {\bibinfo {author} {\bibfnamefont {Y.}~\bibnamefont {Cao}}, \bibinfo {author} {\bibfnamefont {D.}~\bibnamefont {Rodan-Legrain}}, \bibinfo {author} {\bibfnamefont {O.}~\bibnamefont {Rubies-Bigorda}}, \bibinfo {author} {\bibfnamefont {J.~M.}\ \bibnamefont {Park}}, \bibinfo {author} {\bibfnamefont {K.}~\bibnamefont {Watanabe}}, \bibinfo {author} {\bibfnamefont {T.}~\bibnamefont {Taniguchi}}, \ and\ \bibinfo {author} {\bibfnamefont {P.}~\bibnamefont {Jarillo-Herrero}},\ }\href {\doibase 10.1038/s41586-020-2260-6} {\bibfield  {journal} {\bibinfo  {journal} {Nature}\ }\textbf {\bibinfo {volume} {583}},\ \bibinfo {pages} {215} (\bibinfo {year} {2020})}\BibitemShut {NoStop}%
\bibitem [{\citenamefont {Utama}\ \emph {et~al.}(2021{\natexlab{a}})\citenamefont {Utama}, \citenamefont {Koch}, \citenamefont {Lee}, \citenamefont {Leconte}, \citenamefont {Li}, \citenamefont {Zhao}, \citenamefont {Jiang}, \citenamefont {Zhu}, \citenamefont {Watanabe}, \citenamefont {Taniguchi}, \citenamefont {Ashby}, \citenamefont {Weber-Bargioni}, \citenamefont {Zettl}, \citenamefont {Jozwiak}, \citenamefont {Jung}, \citenamefont {Rotenberg}, \citenamefont {Bostwick},\ and\ \citenamefont {Wang}}]{FengWangTBG}%
  \BibitemOpen
  \bibfield  {author} {\bibinfo {author} {\bibfnamefont {M.~I.~B.}\ \bibnamefont {Utama}}, \bibinfo {author} {\bibfnamefont {R.~J.}\ \bibnamefont {Koch}}, \bibinfo {author} {\bibfnamefont {K.}~\bibnamefont {Lee}}, \bibinfo {author} {\bibfnamefont {N.}~\bibnamefont {Leconte}}, \bibinfo {author} {\bibfnamefont {H.}~\bibnamefont {Li}}, \bibinfo {author} {\bibfnamefont {S.}~\bibnamefont {Zhao}}, \bibinfo {author} {\bibfnamefont {L.}~\bibnamefont {Jiang}}, \bibinfo {author} {\bibfnamefont {J.}~\bibnamefont {Zhu}}, \bibinfo {author} {\bibfnamefont {K.}~\bibnamefont {Watanabe}}, \bibinfo {author} {\bibfnamefont {T.}~\bibnamefont {Taniguchi}}, \bibinfo {author} {\bibfnamefont {P.~D.}\ \bibnamefont {Ashby}}, \bibinfo {author} {\bibfnamefont {A.}~\bibnamefont {Weber-Bargioni}}, \bibinfo {author} {\bibfnamefont {A.}~\bibnamefont {Zettl}}, \bibinfo {author} {\bibfnamefont {C.}~\bibnamefont {Jozwiak}}, \bibinfo {author} {\bibfnamefont {J.}~\bibnamefont {Jung}}, \bibinfo {author} {\bibfnamefont {E.}~\bibnamefont
  {Rotenberg}}, \bibinfo {author} {\bibfnamefont {A.}~\bibnamefont {Bostwick}}, \ and\ \bibinfo {author} {\bibfnamefont {F.}~\bibnamefont {Wang}},\ }\href {\doibase 10.1038/s41567-020-0974-x} {\bibfield  {journal} {\bibinfo  {journal} {Nat. Phys.}\ }\textbf {\bibinfo {volume} {17}},\ \bibinfo {pages} {184} (\bibinfo {year} {2021}{\natexlab{a}})}\BibitemShut {NoStop}%
\bibitem [{\citenamefont {Li}\ \emph {et~al.}(2024)\citenamefont {Li}, \citenamefont {Zhang}, \citenamefont {Wang}, \citenamefont {Chen}, \citenamefont {Bao}, \citenamefont {Liu}, \citenamefont {Lin}, \citenamefont {Zhang}, \citenamefont {Zhang}, \citenamefont {Watanabe}, \citenamefont {Taniguchi}, \citenamefont {Avila}, \citenamefont {Dudin}, \citenamefont {Li}, \citenamefont {Yu}, \citenamefont {Duan}, \citenamefont {Song},\ and\ \citenamefont {Zhou}}]{Li2024}%
  \BibitemOpen
  \bibfield  {author} {\bibinfo {author} {\bibfnamefont {Q.}~\bibnamefont {Li}}, \bibinfo {author} {\bibfnamefont {H.}~\bibnamefont {Zhang}}, \bibinfo {author} {\bibfnamefont {Y.}~\bibnamefont {Wang}}, \bibinfo {author} {\bibfnamefont {W.}~\bibnamefont {Chen}}, \bibinfo {author} {\bibfnamefont {C.}~\bibnamefont {Bao}}, \bibinfo {author} {\bibfnamefont {Q.}~\bibnamefont {Liu}}, \bibinfo {author} {\bibfnamefont {T.}~\bibnamefont {Lin}}, \bibinfo {author} {\bibfnamefont {S.}~\bibnamefont {Zhang}}, \bibinfo {author} {\bibfnamefont {H.}~\bibnamefont {Zhang}}, \bibinfo {author} {\bibfnamefont {K.}~\bibnamefont {Watanabe}}, \bibinfo {author} {\bibfnamefont {T.}~\bibnamefont {Taniguchi}}, \bibinfo {author} {\bibfnamefont {J.}~\bibnamefont {Avila}}, \bibinfo {author} {\bibfnamefont {P.}~\bibnamefont {Dudin}}, \bibinfo {author} {\bibfnamefont {Q.}~\bibnamefont {Li}}, \bibinfo {author} {\bibfnamefont {P.}~\bibnamefont {Yu}}, \bibinfo {author} {\bibfnamefont {W.}~\bibnamefont {Duan}}, \bibinfo {author} {\bibfnamefont
  {Z.}~\bibnamefont {Song}}, \ and\ \bibinfo {author} {\bibfnamefont {S.}~\bibnamefont {Zhou}},\ }\href {\doibase 10.1038/s41563-024-01858-4} {\bibfield  {journal} {\bibinfo  {journal} {Nat. Mater.}\ } (\bibinfo {year} {2024}),\ 10.1038/s41563-024-01858-4}\BibitemShut {NoStop}%
\bibitem [{\citenamefont {Yankowitz}\ \emph {et~al.}(2019)\citenamefont {Yankowitz}, \citenamefont {Chen}, \citenamefont {Polshyn}, \citenamefont {Zhang}, \citenamefont {Watanabe}, \citenamefont {Taniguchi}, \citenamefont {Graf}, \citenamefont {Young},\ and\ \citenamefont {Dean}}]{101126scienceaav1910}%
  \BibitemOpen
  \bibfield  {author} {\bibinfo {author} {\bibfnamefont {M.}~\bibnamefont {Yankowitz}}, \bibinfo {author} {\bibfnamefont {S.}~\bibnamefont {Chen}}, \bibinfo {author} {\bibfnamefont {H.}~\bibnamefont {Polshyn}}, \bibinfo {author} {\bibfnamefont {Y.}~\bibnamefont {Zhang}}, \bibinfo {author} {\bibfnamefont {K.}~\bibnamefont {Watanabe}}, \bibinfo {author} {\bibfnamefont {T.}~\bibnamefont {Taniguchi}}, \bibinfo {author} {\bibfnamefont {D.}~\bibnamefont {Graf}}, \bibinfo {author} {\bibfnamefont {A.~F.}\ \bibnamefont {Young}}, \ and\ \bibinfo {author} {\bibfnamefont {C.~R.}\ \bibnamefont {Dean}},\ }\href {\doibase 10.1126/science.aav1910} {\bibfield  {journal} {\bibinfo  {journal} {Science}\ }\textbf {\bibinfo {volume} {363}},\ \bibinfo {pages} {1059} (\bibinfo {year} {2019})}\BibitemShut {NoStop}%
\bibitem [{\citenamefont {Park}\ \emph {et~al.}(2021)\citenamefont {Park}, \citenamefont {Cao}, \citenamefont {Watanabe}, \citenamefont {Taniguchi},\ and\ \citenamefont {Jarillo-Herrero}}]{Park2021}%
  \BibitemOpen
  \bibfield  {author} {\bibinfo {author} {\bibfnamefont {J.~M.}\ \bibnamefont {Park}}, \bibinfo {author} {\bibfnamefont {Y.}~\bibnamefont {Cao}}, \bibinfo {author} {\bibfnamefont {K.}~\bibnamefont {Watanabe}}, \bibinfo {author} {\bibfnamefont {T.}~\bibnamefont {Taniguchi}}, \ and\ \bibinfo {author} {\bibfnamefont {P.}~\bibnamefont {Jarillo-Herrero}},\ }\href {\doibase 10.1038/s41586-021-03192-0} {\bibfield  {journal} {\bibinfo  {journal} {Nature}\ }\textbf {\bibinfo {volume} {590}},\ \bibinfo {pages} {249} (\bibinfo {year} {2021})}\BibitemShut {NoStop}%
\bibitem [{\citenamefont {Codecido}\ \emph {et~al.}(2019)\citenamefont {Codecido}, \citenamefont {Wang}, \citenamefont {Koester}, \citenamefont {Che}, \citenamefont {Tian}, \citenamefont {Lv}, \citenamefont {Tran}, \citenamefont {Watanabe}, \citenamefont {Taniguchi}, \citenamefont {Zhang}, \citenamefont {Bockrath},\ and\ \citenamefont {Lau}}]{101126sciadvaaw9770}%
  \BibitemOpen
  \bibfield  {author} {\bibinfo {author} {\bibfnamefont {E.}~\bibnamefont {Codecido}}, \bibinfo {author} {\bibfnamefont {Q.}~\bibnamefont {Wang}}, \bibinfo {author} {\bibfnamefont {R.}~\bibnamefont {Koester}}, \bibinfo {author} {\bibfnamefont {S.}~\bibnamefont {Che}}, \bibinfo {author} {\bibfnamefont {H.}~\bibnamefont {Tian}}, \bibinfo {author} {\bibfnamefont {R.}~\bibnamefont {Lv}}, \bibinfo {author} {\bibfnamefont {S.}~\bibnamefont {Tran}}, \bibinfo {author} {\bibfnamefont {K.}~\bibnamefont {Watanabe}}, \bibinfo {author} {\bibfnamefont {T.}~\bibnamefont {Taniguchi}}, \bibinfo {author} {\bibfnamefont {F.}~\bibnamefont {Zhang}}, \bibinfo {author} {\bibfnamefont {M.}~\bibnamefont {Bockrath}}, \ and\ \bibinfo {author} {\bibfnamefont {C.~N.}\ \bibnamefont {Lau}},\ }\href {\doibase 10.1126/sciadv.aaw9770} {\bibfield  {journal} {\bibinfo  {journal} {Sci. Adv.}\ }\textbf {\bibinfo {volume} {5}},\ \bibinfo {pages} {eaaw9770} (\bibinfo {year} {2019})}\BibitemShut {NoStop}%
\bibitem [{\citenamefont {Po}\ \emph {et~al.}(2018)\citenamefont {Po}, \citenamefont {Zou}, \citenamefont {Vishwanath},\ and\ \citenamefont {Senthil}}]{PhysRevX.8.031089}%
  \BibitemOpen
  \bibfield  {author} {\bibinfo {author} {\bibfnamefont {H.~C.}\ \bibnamefont {Po}}, \bibinfo {author} {\bibfnamefont {L.}~\bibnamefont {Zou}}, \bibinfo {author} {\bibfnamefont {A.}~\bibnamefont {Vishwanath}}, \ and\ \bibinfo {author} {\bibfnamefont {T.}~\bibnamefont {Senthil}},\ }\href {\doibase 10.1103/PhysRevX.8.031089} {\bibfield  {journal} {\bibinfo  {journal} {Phys. Rev. X}\ }\textbf {\bibinfo {volume} {8}},\ \bibinfo {pages} {031089} (\bibinfo {year} {2018})}\BibitemShut {NoStop}%
\bibitem [{\citenamefont {Shen}\ \emph {et~al.}(2020)\citenamefont {Shen}, \citenamefont {Chu}, \citenamefont {Wu}, \citenamefont {Li}, \citenamefont {Wang}, \citenamefont {Zhao}, \citenamefont {Tang}, \citenamefont {Liu}, \citenamefont {Tian}, \citenamefont {Watanabe}, \citenamefont {Taniguchi}, \citenamefont {Yang}, \citenamefont {Meng}, \citenamefont {Shi}, \citenamefont {Yazyev},\ and\ \citenamefont {Zhang}}]{Shen2020}%
  \BibitemOpen
  \bibfield  {author} {\bibinfo {author} {\bibfnamefont {C.}~\bibnamefont {Shen}}, \bibinfo {author} {\bibfnamefont {Y.}~\bibnamefont {Chu}}, \bibinfo {author} {\bibfnamefont {Q.~S.}\ \bibnamefont {Wu}}, \bibinfo {author} {\bibfnamefont {N.}~\bibnamefont {Li}}, \bibinfo {author} {\bibfnamefont {S.}~\bibnamefont {Wang}}, \bibinfo {author} {\bibfnamefont {Y.}~\bibnamefont {Zhao}}, \bibinfo {author} {\bibfnamefont {J.}~\bibnamefont {Tang}}, \bibinfo {author} {\bibfnamefont {J.}~\bibnamefont {Liu}}, \bibinfo {author} {\bibfnamefont {J.}~\bibnamefont {Tian}}, \bibinfo {author} {\bibfnamefont {K.}~\bibnamefont {Watanabe}}, \bibinfo {author} {\bibfnamefont {T.}~\bibnamefont {Taniguchi}}, \bibinfo {author} {\bibfnamefont {R.}~\bibnamefont {Yang}}, \bibinfo {author} {\bibfnamefont {Z.~Y.}\ \bibnamefont {Meng}}, \bibinfo {author} {\bibfnamefont {D.}~\bibnamefont {Shi}}, \bibinfo {author} {\bibfnamefont {O.~V.}\ \bibnamefont {Yazyev}}, \ and\ \bibinfo {author} {\bibfnamefont {G.}~\bibnamefont {Zhang}},\ }\href {\doibase
  10.1038/s41567-020-0825-9} {\bibfield  {journal} {\bibinfo  {journal} {Nat. Phys.}\ }\textbf {\bibinfo {volume} {16}},\ \bibinfo {pages} {520} (\bibinfo {year} {2020})}\BibitemShut {NoStop}%
\bibitem [{\citenamefont {Lee}\ \emph {et~al.}(2019)\citenamefont {Lee}, \citenamefont {Khalaf}, \citenamefont {Liu}, \citenamefont {Liu}, \citenamefont {Hao}, \citenamefont {Kim},\ and\ \citenamefont {Vishwanath}}]{Lee2019}%
  \BibitemOpen
  \bibfield  {author} {\bibinfo {author} {\bibfnamefont {J.~Y.}\ \bibnamefont {Lee}}, \bibinfo {author} {\bibfnamefont {E.}~\bibnamefont {Khalaf}}, \bibinfo {author} {\bibfnamefont {S.}~\bibnamefont {Liu}}, \bibinfo {author} {\bibfnamefont {X.}~\bibnamefont {Liu}}, \bibinfo {author} {\bibfnamefont {Z.}~\bibnamefont {Hao}}, \bibinfo {author} {\bibfnamefont {P.}~\bibnamefont {Kim}}, \ and\ \bibinfo {author} {\bibfnamefont {A.}~\bibnamefont {Vishwanath}},\ }\href {\doibase 10.1038/s41467-019-12981-1} {\bibfield  {journal} {\bibinfo  {journal} {Nat. Commun.}\ }\textbf {\bibinfo {volume} {10}},\ \bibinfo {pages} {5333} (\bibinfo {year} {2019})}\BibitemShut {NoStop}%
\bibitem [{\citenamefont {Katoch}\ \emph {et~al.}(2018)\citenamefont {Katoch}, \citenamefont {Ulstrup}, \citenamefont {Koch}, \citenamefont {Moser}, \citenamefont {McCreary}, \citenamefont {Singh}, \citenamefont {Xu}, \citenamefont {Jonker}, \citenamefont {Kawakami}, \citenamefont {Bostwick}, \citenamefont {Rotenberg},\ and\ \citenamefont {Jozwiak}}]{Katoch2018}%
  \BibitemOpen
  \bibfield  {author} {\bibinfo {author} {\bibfnamefont {J.}~\bibnamefont {Katoch}}, \bibinfo {author} {\bibfnamefont {S.}~\bibnamefont {Ulstrup}}, \bibinfo {author} {\bibfnamefont {R.~J.}\ \bibnamefont {Koch}}, \bibinfo {author} {\bibfnamefont {S.}~\bibnamefont {Moser}}, \bibinfo {author} {\bibfnamefont {K.~M.}\ \bibnamefont {McCreary}}, \bibinfo {author} {\bibfnamefont {S.}~\bibnamefont {Singh}}, \bibinfo {author} {\bibfnamefont {J.}~\bibnamefont {Xu}}, \bibinfo {author} {\bibfnamefont {B.~T.}\ \bibnamefont {Jonker}}, \bibinfo {author} {\bibfnamefont {R.~K.}\ \bibnamefont {Kawakami}}, \bibinfo {author} {\bibfnamefont {A.}~\bibnamefont {Bostwick}}, \bibinfo {author} {\bibfnamefont {E.}~\bibnamefont {Rotenberg}}, \ and\ \bibinfo {author} {\bibfnamefont {C.}~\bibnamefont {Jozwiak}},\ }\href {\doibase 10.1038/s41567-017-0033-4} {\bibfield  {journal} {\bibinfo  {journal} {Nat. Phys.}\ }\textbf {\bibinfo {volume} {14}},\ \bibinfo {pages} {355} (\bibinfo {year} {2018})}\BibitemShut {NoStop}%
\bibitem [{\citenamefont {Muzzio}\ \emph {et~al.}(2020)\citenamefont {Muzzio}, \citenamefont {Jones}, \citenamefont {Curcio}, \citenamefont {Biswas}, \citenamefont {Miwa}, \citenamefont {Hofmann}, \citenamefont {Watanabe}, \citenamefont {Taniguchi}, \citenamefont {Singh}, \citenamefont {Jozwiak}, \citenamefont {Rotenberg}, \citenamefont {Bostwick}, \citenamefont {Koch}, \citenamefont {Ulstrup},\ and\ \citenamefont {Katoch}}]{RyanPRB}%
  \BibitemOpen
  \bibfield  {author} {\bibinfo {author} {\bibfnamefont {R.}~\bibnamefont {Muzzio}}, \bibinfo {author} {\bibfnamefont {A.~J.~H.}\ \bibnamefont {Jones}}, \bibinfo {author} {\bibfnamefont {D.}~\bibnamefont {Curcio}}, \bibinfo {author} {\bibfnamefont {D.}~\bibnamefont {Biswas}}, \bibinfo {author} {\bibfnamefont {J.~A.}\ \bibnamefont {Miwa}}, \bibinfo {author} {\bibfnamefont {P.}~\bibnamefont {Hofmann}}, \bibinfo {author} {\bibfnamefont {K.}~\bibnamefont {Watanabe}}, \bibinfo {author} {\bibfnamefont {T.}~\bibnamefont {Taniguchi}}, \bibinfo {author} {\bibfnamefont {S.}~\bibnamefont {Singh}}, \bibinfo {author} {\bibfnamefont {C.}~\bibnamefont {Jozwiak}}, \bibinfo {author} {\bibfnamefont {E.}~\bibnamefont {Rotenberg}}, \bibinfo {author} {\bibfnamefont {A.}~\bibnamefont {Bostwick}}, \bibinfo {author} {\bibfnamefont {R.~J.}\ \bibnamefont {Koch}}, \bibinfo {author} {\bibfnamefont {S.}~\bibnamefont {Ulstrup}}, \ and\ \bibinfo {author} {\bibfnamefont {J.}~\bibnamefont {Katoch}},\ }\href {\doibase
  10.1103/PhysRevB.101.201409} {\bibfield  {journal} {\bibinfo  {journal} {Phys. Rev. B}\ }\textbf {\bibinfo {volume} {101}},\ \bibinfo {pages} {201409} (\bibinfo {year} {2020})}\BibitemShut {NoStop}%
\bibitem [{\citenamefont {Koch}\ \emph {et~al.}(2018)\citenamefont {Koch}, \citenamefont {Katoch}, \citenamefont {Moser}, \citenamefont {Schwarz}, \citenamefont {Kawakami}, \citenamefont {Bostwick}, \citenamefont {Rotenberg}, \citenamefont {Jozwiak},\ and\ \citenamefont {Ulstrup}}]{PhysRevMaterials.2.074006}%
  \BibitemOpen
  \bibfield  {author} {\bibinfo {author} {\bibfnamefont {R.~J.}\ \bibnamefont {Koch}}, \bibinfo {author} {\bibfnamefont {J.}~\bibnamefont {Katoch}}, \bibinfo {author} {\bibfnamefont {S.}~\bibnamefont {Moser}}, \bibinfo {author} {\bibfnamefont {D.}~\bibnamefont {Schwarz}}, \bibinfo {author} {\bibfnamefont {R.~K.}\ \bibnamefont {Kawakami}}, \bibinfo {author} {\bibfnamefont {A.}~\bibnamefont {Bostwick}}, \bibinfo {author} {\bibfnamefont {E.}~\bibnamefont {Rotenberg}}, \bibinfo {author} {\bibfnamefont {C.}~\bibnamefont {Jozwiak}}, \ and\ \bibinfo {author} {\bibfnamefont {S.}~\bibnamefont {Ulstrup}},\ }\href {\doibase 10.1103/PhysRevMaterials.2.074006} {\bibfield  {journal} {\bibinfo  {journal} {Phys. Rev. Mater.}\ }\textbf {\bibinfo {volume} {2}},\ \bibinfo {pages} {074006} (\bibinfo {year} {2018})}\BibitemShut {NoStop}%
\bibitem [{\citenamefont {Jones}\ \emph {et~al.}(2020)\citenamefont {Jones}, \citenamefont {Muzzio}, \citenamefont {Majchrzak}, \citenamefont {Pakdel}, \citenamefont {Curcio}, \citenamefont {Volckaert}, \citenamefont {Biswas}, \citenamefont {Gobbo}, \citenamefont {Singh}, \citenamefont {Robinson}, \citenamefont {Watanabe}, \citenamefont {Taniguchi}, \citenamefont {Kim}, \citenamefont {Cacho}, \citenamefont {Lanata}, \citenamefont {Miwa}, \citenamefont {Hofmann}, \citenamefont {Katoch},\ and\ \citenamefont {Ulstrup}}]{AlfredAdvaceMaterial}%
  \BibitemOpen
  \bibfield  {author} {\bibinfo {author} {\bibfnamefont {A.~J.~H.}\ \bibnamefont {Jones}}, \bibinfo {author} {\bibfnamefont {R.}~\bibnamefont {Muzzio}}, \bibinfo {author} {\bibfnamefont {P.}~\bibnamefont {Majchrzak}}, \bibinfo {author} {\bibfnamefont {S.}~\bibnamefont {Pakdel}}, \bibinfo {author} {\bibfnamefont {D.}~\bibnamefont {Curcio}}, \bibinfo {author} {\bibfnamefont {K.}~\bibnamefont {Volckaert}}, \bibinfo {author} {\bibfnamefont {D.}~\bibnamefont {Biswas}}, \bibinfo {author} {\bibfnamefont {J.}~\bibnamefont {Gobbo}}, \bibinfo {author} {\bibfnamefont {S.}~\bibnamefont {Singh}}, \bibinfo {author} {\bibfnamefont {J.~T.}\ \bibnamefont {Robinson}}, \bibinfo {author} {\bibfnamefont {K.}~\bibnamefont {Watanabe}}, \bibinfo {author} {\bibfnamefont {T.}~\bibnamefont {Taniguchi}}, \bibinfo {author} {\bibfnamefont {T.~K.}\ \bibnamefont {Kim}}, \bibinfo {author} {\bibfnamefont {C.}~\bibnamefont {Cacho}}, \bibinfo {author} {\bibfnamefont {N.}~\bibnamefont {Lanata}}, \bibinfo {author} {\bibfnamefont {J.~A.}\
  \bibnamefont {Miwa}}, \bibinfo {author} {\bibfnamefont {P.}~\bibnamefont {Hofmann}}, \bibinfo {author} {\bibfnamefont {J.}~\bibnamefont {Katoch}}, \ and\ \bibinfo {author} {\bibfnamefont {S.}~\bibnamefont {Ulstrup}},\ }\href {\doibase https://doi.org/10.1002/adma.202001656} {\bibfield  {journal} {\bibinfo  {journal} {Adv. Mater.}\ }\textbf {\bibinfo {volume} {32}},\ \bibinfo {pages} {2001656} (\bibinfo {year} {2020})}\BibitemShut {NoStop}%
\bibitem [{\citenamefont {Chen}\ \emph {et~al.}(2024)\citenamefont {Chen}, \citenamefont {Nuckolls}, \citenamefont {Ding}, \citenamefont {Miao}, \citenamefont {Wong}, \citenamefont {Oh}, \citenamefont {Lee}, \citenamefont {He}, \citenamefont {Peng}, \citenamefont {Pei}, \citenamefont {Li}, \citenamefont {Hao}, \citenamefont {Yan}, \citenamefont {Xiao}, \citenamefont {Gao}, \citenamefont {Li}, \citenamefont {Zhang}, \citenamefont {Liu}, \citenamefont {He}, \citenamefont {Watanabe}, \citenamefont {Taniguchi}, \citenamefont {Jozwiak}, \citenamefont {Bostwick}, \citenamefont {Rotenberg}, \citenamefont {Li}, \citenamefont {Han}, \citenamefont {Pan}, \citenamefont {Liu}, \citenamefont {Dai}, \citenamefont {Liu}, \citenamefont {Bernevig}, \citenamefont {Wang}, \citenamefont {Yazdani},\ and\ \citenamefont {Chen}}]{Chen2024}%
  \BibitemOpen
  \bibfield  {author} {\bibinfo {author} {\bibfnamefont {C.}~\bibnamefont {Chen}}, \bibinfo {author} {\bibfnamefont {K.~P.}\ \bibnamefont {Nuckolls}}, \bibinfo {author} {\bibfnamefont {S.}~\bibnamefont {Ding}}, \bibinfo {author} {\bibfnamefont {W.}~\bibnamefont {Miao}}, \bibinfo {author} {\bibfnamefont {D.}~\bibnamefont {Wong}}, \bibinfo {author} {\bibfnamefont {M.}~\bibnamefont {Oh}}, \bibinfo {author} {\bibfnamefont {R.~L.}\ \bibnamefont {Lee}}, \bibinfo {author} {\bibfnamefont {S.}~\bibnamefont {He}}, \bibinfo {author} {\bibfnamefont {C.}~\bibnamefont {Peng}}, \bibinfo {author} {\bibfnamefont {D.}~\bibnamefont {Pei}}, \bibinfo {author} {\bibfnamefont {Y.}~\bibnamefont {Li}}, \bibinfo {author} {\bibfnamefont {C.}~\bibnamefont {Hao}}, \bibinfo {author} {\bibfnamefont {H.}~\bibnamefont {Yan}}, \bibinfo {author} {\bibfnamefont {H.}~\bibnamefont {Xiao}}, \bibinfo {author} {\bibfnamefont {H.}~\bibnamefont {Gao}}, \bibinfo {author} {\bibfnamefont {Q.}~\bibnamefont {Li}}, \bibinfo {author} {\bibfnamefont
  {S.}~\bibnamefont {Zhang}}, \bibinfo {author} {\bibfnamefont {J.}~\bibnamefont {Liu}}, \bibinfo {author} {\bibfnamefont {L.}~\bibnamefont {He}}, \bibinfo {author} {\bibfnamefont {K.}~\bibnamefont {Watanabe}}, \bibinfo {author} {\bibfnamefont {T.}~\bibnamefont {Taniguchi}}, \bibinfo {author} {\bibfnamefont {C.}~\bibnamefont {Jozwiak}}, \bibinfo {author} {\bibfnamefont {A.}~\bibnamefont {Bostwick}}, \bibinfo {author} {\bibfnamefont {E.}~\bibnamefont {Rotenberg}}, \bibinfo {author} {\bibfnamefont {C.}~\bibnamefont {Li}}, \bibinfo {author} {\bibfnamefont {X.}~\bibnamefont {Han}}, \bibinfo {author} {\bibfnamefont {D.}~\bibnamefont {Pan}}, \bibinfo {author} {\bibfnamefont {Z.}~\bibnamefont {Liu}}, \bibinfo {author} {\bibfnamefont {X.}~\bibnamefont {Dai}}, \bibinfo {author} {\bibfnamefont {C.}~\bibnamefont {Liu}}, \bibinfo {author} {\bibfnamefont {B.~A.}\ \bibnamefont {Bernevig}}, \bibinfo {author} {\bibfnamefont {Y.}~\bibnamefont {Wang}}, \bibinfo {author} {\bibfnamefont {A.}~\bibnamefont {Yazdani}}, \ and\
  \bibinfo {author} {\bibfnamefont {Y.}~\bibnamefont {Chen}},\ }\href {\doibase 10.1038/s41586-024-08227-w} {\bibfield  {journal} {\bibinfo  {journal} {Nature}\ }\textbf {\bibinfo {volume} {636}},\ \bibinfo {pages} {342} (\bibinfo {year} {2024})}\BibitemShut {NoStop}%
\bibitem [{sup()}]{supply}%
  \BibitemOpen
  \href@noop {} {\bibinfo  {journal} {See Supplemental Materials at \url{https://doi.org/10.1103/3vg5-vkzy} for sample Fabrication, twist angle, \rm{AFM}, MDCs cut, which includes Ref.~\cite{science1130681,PhysRevB.51.6868,httpsdoi.org10.1002smll.202300144,Trolle2017, PhysRevMaterials.7.054001}}\ }\BibitemShut {NoStop}%
\bibitem [{\citenamefont {Joucken}\ \emph {et~al.}(2019)\citenamefont {Joucken}, \citenamefont {Henrard},\ and\ \citenamefont {Lagoute}}]{PhysRevMaterials.3.110301}%
  \BibitemOpen
\bibfield  {journal} {  }\bibfield  {author} {\bibinfo {author} {\bibfnamefont {F.}~\bibnamefont {Joucken}}, \bibinfo {author} {\bibfnamefont {L.}~\bibnamefont {Henrard}}, \ and\ \bibinfo {author} {\bibfnamefont {J.}~\bibnamefont {Lagoute}},\ }\href {\doibase 10.1103/PhysRevMaterials.3.110301} {\bibfield  {journal} {\bibinfo  {journal} {Phys. Rev. Mater.}\ }\textbf {\bibinfo {volume} {3}},\ \bibinfo {pages} {110301} (\bibinfo {year} {2019})}\BibitemShut {NoStop}%
\bibitem [{\citenamefont {Singh}\ \emph {et~al.}(2023)\citenamefont {Singh}, \citenamefont {Andleeb},\ and\ \citenamefont {Singh}}]{10.1063/5.0164903}%
  \BibitemOpen
  \bibfield  {author} {\bibinfo {author} {\bibfnamefont {A.~K.}\ \bibnamefont {Singh}}, \bibinfo {author} {\bibfnamefont {S.}~\bibnamefont {Andleeb}}, \ and\ \bibinfo {author} {\bibfnamefont {A.~K.}\ \bibnamefont {Singh}},\ }\href {\doibase 10.1063/5.0164903} {\bibfield  {journal} {\bibinfo  {journal} {AIP Advances}\ }\textbf {\bibinfo {volume} {13}},\ \bibinfo {pages} {095012} (\bibinfo {year} {2023})}\BibitemShut {NoStop}%
\bibitem [{\citenamefont {Nezval}\ \emph {et~al.}(2023)\citenamefont {Nezval}, \citenamefont {Bartošík}, \citenamefont {Mach}, \citenamefont {Švarc}, \citenamefont {Konečný}, \citenamefont {Piastek}, \citenamefont {Špaček},\ and\ \citenamefont {Šikola}}]{10.1063/5.0161160}%
  \BibitemOpen
  \bibfield  {author} {\bibinfo {author} {\bibfnamefont {D.}~\bibnamefont {Nezval}}, \bibinfo {author} {\bibfnamefont {M.}~\bibnamefont {Bartošík}}, \bibinfo {author} {\bibfnamefont {J.}~\bibnamefont {Mach}}, \bibinfo {author} {\bibfnamefont {V.}~\bibnamefont {Švarc}}, \bibinfo {author} {\bibfnamefont {M.}~\bibnamefont {Konečný}}, \bibinfo {author} {\bibfnamefont {J.}~\bibnamefont {Piastek}}, \bibinfo {author} {\bibfnamefont {O.}~\bibnamefont {Špaček}}, \ and\ \bibinfo {author} {\bibfnamefont {T.}~\bibnamefont {Šikola}},\ }\href {\doibase 10.1063/5.0161160} {\bibfield  {journal} {\bibinfo  {journal} {The Journal of Chemical Physics}\ }\textbf {\bibinfo {volume} {159}},\ \bibinfo {pages} {214710} (\bibinfo {year} {2023})}\BibitemShut {NoStop}%
\bibitem [{\citenamefont {Giovannetti}\ \emph {et~al.}(2008)\citenamefont {Giovannetti}, \citenamefont {Khomyakov}, \citenamefont {Brocks}, \citenamefont {Karpan}, \citenamefont {van~den Brink},\ and\ \citenamefont {Kelly}}]{PhysRevLett.101.026803}%
  \BibitemOpen
  \bibfield  {author} {\bibinfo {author} {\bibfnamefont {G.}~\bibnamefont {Giovannetti}}, \bibinfo {author} {\bibfnamefont {P.~A.}\ \bibnamefont {Khomyakov}}, \bibinfo {author} {\bibfnamefont {G.}~\bibnamefont {Brocks}}, \bibinfo {author} {\bibfnamefont {V.~M.}\ \bibnamefont {Karpan}}, \bibinfo {author} {\bibfnamefont {J.}~\bibnamefont {van~den Brink}}, \ and\ \bibinfo {author} {\bibfnamefont {P.~J.}\ \bibnamefont {Kelly}},\ }\href {\doibase 10.1103/PhysRevLett.101.026803} {\bibfield  {journal} {\bibinfo  {journal} {Phys. Rev. Lett.}\ }\textbf {\bibinfo {volume} {101}},\ \bibinfo {pages} {026803} (\bibinfo {year} {2008})}\BibitemShut {NoStop}%
\bibitem [{\citenamefont {McCann}(2011)}]{McCann_2011}%
  \BibitemOpen
  \bibfield  {author} {\bibinfo {author} {\bibfnamefont {E.}~\bibnamefont {McCann}},\ }\enquote {\bibinfo {title} {Electronic properties of monolayer and bilayer graphene},}\ in\ \href {\doibase 10.1007/978-3-642-22984-8_8} {\emph {\bibinfo {booktitle} {Graphene Nanoelectronics}}}\ (\bibinfo  {publisher} {Springer Berlin Heidelberg},\ \bibinfo {year} {2011})\ p.\ \bibinfo {pages} {237–275}\BibitemShut {NoStop}%
\bibitem [{\citenamefont {Raza}(2012)}]{raza2012graphene}%
  \BibitemOpen
  \bibfield  {author} {\bibinfo {author} {\bibfnamefont {H.}~\bibnamefont {Raza}},\ }\href@noop {} {\emph {\bibinfo {title} {Graphene nanoelectronics: Metrology, synthesis, properties and applications}}}\ (\bibinfo  {publisher} {Springer Science \& Business Media},\ \bibinfo {year} {2012})\BibitemShut {NoStop}%
\bibitem [{\citenamefont {Nunn}\ \emph {et~al.}(2023)\citenamefont {Nunn}, \citenamefont {McEllistrim}, \citenamefont {Weston}, \citenamefont {Garcia-Ruiz}, \citenamefont {Watson}, \citenamefont {Mucha-Kruczynski}, \citenamefont {Cacho}, \citenamefont {Gorbachev}, \citenamefont {Fal'ko},\ and\ \citenamefont {Wilson}}]{WilsonTG}%
  \BibitemOpen
  \bibfield  {author} {\bibinfo {author} {\bibfnamefont {J.~E.}\ \bibnamefont {Nunn}}, \bibinfo {author} {\bibfnamefont {A.}~\bibnamefont {McEllistrim}}, \bibinfo {author} {\bibfnamefont {A.}~\bibnamefont {Weston}}, \bibinfo {author} {\bibfnamefont {A.}~\bibnamefont {Garcia-Ruiz}}, \bibinfo {author} {\bibfnamefont {M.~D.}\ \bibnamefont {Watson}}, \bibinfo {author} {\bibfnamefont {M.}~\bibnamefont {Mucha-Kruczynski}}, \bibinfo {author} {\bibfnamefont {C.}~\bibnamefont {Cacho}}, \bibinfo {author} {\bibfnamefont {R.~V.}\ \bibnamefont {Gorbachev}}, \bibinfo {author} {\bibfnamefont {V.~I.}\ \bibnamefont {Fal'ko}}, \ and\ \bibinfo {author} {\bibfnamefont {N.~R.}\ \bibnamefont {Wilson}},\ }\bibfield  {booktitle} {\emph {\bibinfo {booktitle} {Nano Letters}},\ }\href {\doibase 10.1021/acs.nanolett.3c01173} {\bibfield  {journal} {\bibinfo  {journal} {Nano Letters}\ }\textbf {\bibinfo {volume} {23}},\ \bibinfo {pages} {5201} (\bibinfo {year} {2023})}\BibitemShut {NoStop}%
\bibitem [{\citenamefont {He}\ \emph {et~al.}(2021)\citenamefont {He}, \citenamefont {Li}, \citenamefont {Cai}, \citenamefont {Liu}, \citenamefont {Watanabe}, \citenamefont {Taniguchi}, \citenamefont {Xu},\ and\ \citenamefont {Yankowitz}}]{He2021}%
  \BibitemOpen
  \bibfield  {author} {\bibinfo {author} {\bibfnamefont {M.}~\bibnamefont {He}}, \bibinfo {author} {\bibfnamefont {Y.}~\bibnamefont {Li}}, \bibinfo {author} {\bibfnamefont {J.}~\bibnamefont {Cai}}, \bibinfo {author} {\bibfnamefont {Y.}~\bibnamefont {Liu}}, \bibinfo {author} {\bibfnamefont {K.}~\bibnamefont {Watanabe}}, \bibinfo {author} {\bibfnamefont {T.}~\bibnamefont {Taniguchi}}, \bibinfo {author} {\bibfnamefont {X.}~\bibnamefont {Xu}}, \ and\ \bibinfo {author} {\bibfnamefont {M.}~\bibnamefont {Yankowitz}},\ }\href {\doibase 10.1038/s41567-020-1030-6} {\bibfield  {journal} {\bibinfo  {journal} {Nat. Phys.}\ }\textbf {\bibinfo {volume} {17}},\ \bibinfo {pages} {26} (\bibinfo {year} {2021})}\BibitemShut {NoStop}%
\bibitem [{\citenamefont {Su}\ \emph {et~al.}(2023)\citenamefont {Su}, \citenamefont {Kuiri}, \citenamefont {Watanabe}, \citenamefont {Taniguchi},\ and\ \citenamefont {Folk}}]{Su2023}%
  \BibitemOpen
  \bibfield  {author} {\bibinfo {author} {\bibfnamefont {R.}~\bibnamefont {Su}}, \bibinfo {author} {\bibfnamefont {M.}~\bibnamefont {Kuiri}}, \bibinfo {author} {\bibfnamefont {K.}~\bibnamefont {Watanabe}}, \bibinfo {author} {\bibfnamefont {T.}~\bibnamefont {Taniguchi}}, \ and\ \bibinfo {author} {\bibfnamefont {J.}~\bibnamefont {Folk}},\ }\href {\doibase 10.1038/s41563-023-01653-7} {\bibfield  {journal} {\bibinfo  {journal} {Nat. Mater.}\ }\textbf {\bibinfo {volume} {22}},\ \bibinfo {pages} {1332} (\bibinfo {year} {2023})}\BibitemShut {NoStop}%
\bibitem [{\citenamefont {Liu}\ \emph {et~al.}(2021)\citenamefont {Liu}, \citenamefont {Chiu}, \citenamefont {Lee}, \citenamefont {Farahi}, \citenamefont {Watanabe}, \citenamefont {Taniguchi}, \citenamefont {Vishwanath},\ and\ \citenamefont {Yazdani}}]{Liu2021}%
  \BibitemOpen
  \bibfield  {author} {\bibinfo {author} {\bibfnamefont {X.}~\bibnamefont {Liu}}, \bibinfo {author} {\bibfnamefont {C.-L.}\ \bibnamefont {Chiu}}, \bibinfo {author} {\bibfnamefont {J.~Y.}\ \bibnamefont {Lee}}, \bibinfo {author} {\bibfnamefont {G.}~\bibnamefont {Farahi}}, \bibinfo {author} {\bibfnamefont {K.}~\bibnamefont {Watanabe}}, \bibinfo {author} {\bibfnamefont {T.}~\bibnamefont {Taniguchi}}, \bibinfo {author} {\bibfnamefont {A.}~\bibnamefont {Vishwanath}}, \ and\ \bibinfo {author} {\bibfnamefont {A.}~\bibnamefont {Yazdani}},\ }\href {\doibase 10.1038/s41467-021-23031-0} {\bibfield  {journal} {\bibinfo  {journal} {Nat. Commun.}\ }\textbf {\bibinfo {volume} {12}},\ \bibinfo {pages} {2732} (\bibinfo {year} {2021})}\BibitemShut {NoStop}%
\bibitem [{\citenamefont {Haddadi}\ \emph {et~al.}(2020)\citenamefont {Haddadi}, \citenamefont {Wu}, \citenamefont {Kruchkov},\ and\ \citenamefont {Yazyev}}]{Haddadi2020}%
  \BibitemOpen
  \bibfield  {author} {\bibinfo {author} {\bibfnamefont {F.}~\bibnamefont {Haddadi}}, \bibinfo {author} {\bibfnamefont {Q.}~\bibnamefont {Wu}}, \bibinfo {author} {\bibfnamefont {A.~J.}\ \bibnamefont {Kruchkov}}, \ and\ \bibinfo {author} {\bibfnamefont {O.~V.}\ \bibnamefont {Yazyev}},\ }\href {\doibase 10.1021/acs.nanolett.9b05117} {\bibfield  {journal} {\bibinfo  {journal} {Nano Letters}\ }\textbf {\bibinfo {volume} {20}},\ \bibinfo {pages} {2410} (\bibinfo {year} {2020})}\BibitemShut {NoStop}%
\bibitem [{\citenamefont {Utama}\ \emph {et~al.}(2021{\natexlab{b}})\citenamefont {Utama}, \citenamefont {Koch}, \citenamefont {Lee}, \citenamefont {Leconte}, \citenamefont {Li}, \citenamefont {Zhao}, \citenamefont {Jiang}, \citenamefont {Zhu}, \citenamefont {Watanabe}, \citenamefont {Taniguchi}, \citenamefont {Ashby}, \citenamefont {Weber-Bargioni}, \citenamefont {Zettl}, \citenamefont {Jozwiak}, \citenamefont {Jung}, \citenamefont {Rotenberg}, \citenamefont {Bostwick},\ and\ \citenamefont {Wang}}]{Utama2021}%
  \BibitemOpen
  \bibfield  {author} {\bibinfo {author} {\bibfnamefont {M.~I.~B.}\ \bibnamefont {Utama}}, \bibinfo {author} {\bibfnamefont {R.~J.}\ \bibnamefont {Koch}}, \bibinfo {author} {\bibfnamefont {K.}~\bibnamefont {Lee}}, \bibinfo {author} {\bibfnamefont {N.}~\bibnamefont {Leconte}}, \bibinfo {author} {\bibfnamefont {H.}~\bibnamefont {Li}}, \bibinfo {author} {\bibfnamefont {S.}~\bibnamefont {Zhao}}, \bibinfo {author} {\bibfnamefont {L.}~\bibnamefont {Jiang}}, \bibinfo {author} {\bibfnamefont {J.}~\bibnamefont {Zhu}}, \bibinfo {author} {\bibfnamefont {K.}~\bibnamefont {Watanabe}}, \bibinfo {author} {\bibfnamefont {T.}~\bibnamefont {Taniguchi}}, \bibinfo {author} {\bibfnamefont {P.~D.}\ \bibnamefont {Ashby}}, \bibinfo {author} {\bibfnamefont {A.}~\bibnamefont {Weber-Bargioni}}, \bibinfo {author} {\bibfnamefont {A.}~\bibnamefont {Zettl}}, \bibinfo {author} {\bibfnamefont {C.}~\bibnamefont {Jozwiak}}, \bibinfo {author} {\bibfnamefont {J.}~\bibnamefont {Jung}}, \bibinfo {author} {\bibfnamefont {E.}~\bibnamefont
  {Rotenberg}}, \bibinfo {author} {\bibfnamefont {A.}~\bibnamefont {Bostwick}}, \ and\ \bibinfo {author} {\bibfnamefont {F.}~\bibnamefont {Wang}},\ }\href {\doibase 10.1038/s41567-020-0974-x} {\bibfield  {journal} {\bibinfo  {journal} {Nat. Phys.}\ }\textbf {\bibinfo {volume} {17}},\ \bibinfo {pages} {184} (\bibinfo {year} {2021}{\natexlab{b}})}\BibitemShut {NoStop}%
\bibitem [{\citenamefont {Adak}\ \emph {et~al.}(2020)\citenamefont {Adak}, \citenamefont {Sinha}, \citenamefont {Ghorai}, \citenamefont {Sangani}, \citenamefont {Watanabe}, \citenamefont {Taniguchi}, \citenamefont {Sensarma},\ and\ \citenamefont {Deshmukh}}]{PhysRevB.101.125428}%
  \BibitemOpen
  \bibfield  {author} {\bibinfo {author} {\bibfnamefont {P.~C.}\ \bibnamefont {Adak}}, \bibinfo {author} {\bibfnamefont {S.}~\bibnamefont {Sinha}}, \bibinfo {author} {\bibfnamefont {U.}~\bibnamefont {Ghorai}}, \bibinfo {author} {\bibfnamefont {L.~D.~V.}\ \bibnamefont {Sangani}}, \bibinfo {author} {\bibfnamefont {K.}~\bibnamefont {Watanabe}}, \bibinfo {author} {\bibfnamefont {T.}~\bibnamefont {Taniguchi}}, \bibinfo {author} {\bibfnamefont {R.}~\bibnamefont {Sensarma}}, \ and\ \bibinfo {author} {\bibfnamefont {M.~M.}\ \bibnamefont {Deshmukh}},\ }\href {\doibase 10.1103/PhysRevB.101.125428} {\bibfield  {journal} {\bibinfo  {journal} {Phys. Rev. B}\ }\textbf {\bibinfo {volume} {101}},\ \bibinfo {pages} {125428} (\bibinfo {year} {2020})}\BibitemShut {NoStop}%
\bibitem [{\citenamefont {Yang}\ \emph {et~al.}(2022)\citenamefont {Yang}, \citenamefont {Chen}, \citenamefont {Han}, \citenamefont {Zhang}, \citenamefont {Zhang}, \citenamefont {Jiang}, \citenamefont {Lyu}, \citenamefont {Li}, \citenamefont {Watanabe}, \citenamefont {Taniguchi}, \citenamefont {Shi}, \citenamefont {Senthil}, \citenamefont {Zhang}, \citenamefont {Wang},\ and\ \citenamefont {Ju}}]{101126scienceabg3036}%
  \BibitemOpen
  \bibfield  {author} {\bibinfo {author} {\bibfnamefont {J.}~\bibnamefont {Yang}}, \bibinfo {author} {\bibfnamefont {G.}~\bibnamefont {Chen}}, \bibinfo {author} {\bibfnamefont {T.}~\bibnamefont {Han}}, \bibinfo {author} {\bibfnamefont {Q.}~\bibnamefont {Zhang}}, \bibinfo {author} {\bibfnamefont {Y.-H.}\ \bibnamefont {Zhang}}, \bibinfo {author} {\bibfnamefont {L.}~\bibnamefont {Jiang}}, \bibinfo {author} {\bibfnamefont {B.}~\bibnamefont {Lyu}}, \bibinfo {author} {\bibfnamefont {H.}~\bibnamefont {Li}}, \bibinfo {author} {\bibfnamefont {K.}~\bibnamefont {Watanabe}}, \bibinfo {author} {\bibfnamefont {T.}~\bibnamefont {Taniguchi}}, \bibinfo {author} {\bibfnamefont {Z.}~\bibnamefont {Shi}}, \bibinfo {author} {\bibfnamefont {T.}~\bibnamefont {Senthil}}, \bibinfo {author} {\bibfnamefont {Y.}~\bibnamefont {Zhang}}, \bibinfo {author} {\bibfnamefont {F.}~\bibnamefont {Wang}}, \ and\ \bibinfo {author} {\bibfnamefont {L.}~\bibnamefont {Ju}},\ }\href {\doibase 10.1126/science.abg3036} {\bibfield  {journal} {\bibinfo
  {journal} {Science}\ }\textbf {\bibinfo {volume} {375}},\ \bibinfo {pages} {1295} (\bibinfo {year} {2022})}\BibitemShut {NoStop}%
\bibitem [{\citenamefont {Jiang}\ \emph {et~al.}(2025)\citenamefont {Jiang}, \citenamefont {Lee}, \citenamefont {Jones}, \citenamefont {Park}, \citenamefont {Hsieh}, \citenamefont {Majchrzak}, \citenamefont {Sahoo}, \citenamefont {Nielsen}, \citenamefont {Watanabe}, \citenamefont {Taniguchi}, \citenamefont {Hofmann}, \citenamefont {Miwa}, \citenamefont {Chen}, \citenamefont {Jung},\ and\ \citenamefont {Ulstrup}}]{Jiang2025}%
  \BibitemOpen
  \bibfield  {author} {\bibinfo {author} {\bibfnamefont {Z.}~\bibnamefont {Jiang}}, \bibinfo {author} {\bibfnamefont {D.}~\bibnamefont {Lee}}, \bibinfo {author} {\bibfnamefont {A.~J.~H.}\ \bibnamefont {Jones}}, \bibinfo {author} {\bibfnamefont {Y.}~\bibnamefont {Park}}, \bibinfo {author} {\bibfnamefont {K.}~\bibnamefont {Hsieh}}, \bibinfo {author} {\bibfnamefont {P.}~\bibnamefont {Majchrzak}}, \bibinfo {author} {\bibfnamefont {C.}~\bibnamefont {Sahoo}}, \bibinfo {author} {\bibfnamefont {T.~S.}\ \bibnamefont {Nielsen}}, \bibinfo {author} {\bibfnamefont {K.}~\bibnamefont {Watanabe}}, \bibinfo {author} {\bibfnamefont {T.}~\bibnamefont {Taniguchi}}, \bibinfo {author} {\bibfnamefont {P.}~\bibnamefont {Hofmann}}, \bibinfo {author} {\bibfnamefont {J.~A.}\ \bibnamefont {Miwa}}, \bibinfo {author} {\bibfnamefont {Y.~P.}\ \bibnamefont {Chen}}, \bibinfo {author} {\bibfnamefont {J.}~\bibnamefont {Jung}}, \ and\ \bibinfo {author} {\bibfnamefont {S.}~\bibnamefont {Ulstrup}},\ }\href {\doibase 10.1021/acsnano.4c12905}
  {\bibfield  {journal} {\bibinfo  {journal} {ACS Nano}\ }\textbf {\bibinfo {volume} {19}},\ \bibinfo {pages} {2379} (\bibinfo {year} {2025})}\BibitemShut {NoStop}%
\bibitem [{\citenamefont {Wang}\ \emph {et~al.}(2024)\citenamefont {Wang}, \citenamefont {Burg}, \citenamefont {Lian}, \citenamefont {Watanabe}, \citenamefont {Taniguchi}, \citenamefont {Bernevig},\ and\ \citenamefont {Tutuc}}]{PhysRevLett.133.246401}%
  \BibitemOpen
  \bibfield  {author} {\bibinfo {author} {\bibfnamefont {Y.}~\bibnamefont {Wang}}, \bibinfo {author} {\bibfnamefont {G.~W.}\ \bibnamefont {Burg}}, \bibinfo {author} {\bibfnamefont {B.}~\bibnamefont {Lian}}, \bibinfo {author} {\bibfnamefont {K.}~\bibnamefont {Watanabe}}, \bibinfo {author} {\bibfnamefont {T.}~\bibnamefont {Taniguchi}}, \bibinfo {author} {\bibfnamefont {B.~A.}\ \bibnamefont {Bernevig}}, \ and\ \bibinfo {author} {\bibfnamefont {E.}~\bibnamefont {Tutuc}},\ }\href {\doibase 10.1103/PhysRevLett.133.246401} {\bibfield  {journal} {\bibinfo  {journal} {Phys. Rev. Lett.}\ }\textbf {\bibinfo {volume} {133}},\ \bibinfo {pages} {246401} (\bibinfo {year} {2024})}\BibitemShut {NoStop}%
\bibitem [{\citenamefont {Zhu}\ \emph {et~al.}(2025)\citenamefont {Zhu}, \citenamefont {Liu}, \citenamefont {Yuan}, \citenamefont {Lu}, \citenamefont {Dong}, \citenamefont {Chu}, \citenamefont {Du}, \citenamefont {Watanabe}, \citenamefont {Taniguchi}, \citenamefont {Shi}, \citenamefont {Wu}, \citenamefont {Liu}, \citenamefont {Zhang},\ and\ \citenamefont {Yang}}]{qc2g-qjg2}%
  \BibitemOpen
  \bibfield  {author} {\bibinfo {author} {\bibfnamefont {J.}~\bibnamefont {Zhu}}, \bibinfo {author} {\bibfnamefont {L.}~\bibnamefont {Liu}}, \bibinfo {author} {\bibfnamefont {Y.}~\bibnamefont {Yuan}}, \bibinfo {author} {\bibfnamefont {X.}~\bibnamefont {Lu}}, \bibinfo {author} {\bibfnamefont {J.}~\bibnamefont {Dong}}, \bibinfo {author} {\bibfnamefont {Y.}~\bibnamefont {Chu}}, \bibinfo {author} {\bibfnamefont {L.}~\bibnamefont {Du}}, \bibinfo {author} {\bibfnamefont {K.}~\bibnamefont {Watanabe}}, \bibinfo {author} {\bibfnamefont {T.}~\bibnamefont {Taniguchi}}, \bibinfo {author} {\bibfnamefont {D.}~\bibnamefont {Shi}}, \bibinfo {author} {\bibfnamefont {Q.}~\bibnamefont {Wu}}, \bibinfo {author} {\bibfnamefont {J.}~\bibnamefont {Liu}}, \bibinfo {author} {\bibfnamefont {G.}~\bibnamefont {Zhang}}, \ and\ \bibinfo {author} {\bibfnamefont {W.}~\bibnamefont {Yang}},\ }\href {\doibase 10.1103/qc2g-qjg2} {\bibfield  {journal} {\bibinfo  {journal} {Phys. Rev. B}\ }\textbf {\bibinfo {volume} {112}},\ \bibinfo {pages}
  {L081108} (\bibinfo {year} {2025})}\BibitemShut {NoStop}%
\bibitem [{\citenamefont {Koshino}(2019)}]{PhysRevB.99.235406}%
  \BibitemOpen
  \bibfield  {author} {\bibinfo {author} {\bibfnamefont {M.}~\bibnamefont {Koshino}},\ }\href {\doibase 10.1103/PhysRevB.99.235406} {\bibfield  {journal} {\bibinfo  {journal} {Phys. Rev. B}\ }\textbf {\bibinfo {volume} {99}},\ \bibinfo {pages} {235406} (\bibinfo {year} {2019})}\BibitemShut {NoStop}%
\bibitem [{\citenamefont {Perea-Causin}\ \emph {et~al.}(2025)\citenamefont {Perea-Causin}, \citenamefont {Liu},\ and\ \citenamefont {Bergholtz}}]{Perea-Causin2025}%
  \BibitemOpen
  \bibfield  {author} {\bibinfo {author} {\bibfnamefont {R.}~\bibnamefont {Perea-Causin}}, \bibinfo {author} {\bibfnamefont {H.}~\bibnamefont {Liu}}, \ and\ \bibinfo {author} {\bibfnamefont {E.~J.}\ \bibnamefont {Bergholtz}},\ }\href {\doibase 10.1038/s41467-025-62224-9} {\bibfield  {journal} {\bibinfo  {journal} {Nat. Commun.}\ }\textbf {\bibinfo {volume} {16}},\ \bibinfo {pages} {6875} (\bibinfo {year} {2025})}\BibitemShut {NoStop}%
\bibitem [{\citenamefont {Ohta}\ \emph {et~al.}(2006)\citenamefont {Ohta}, \citenamefont {Bostwick}, \citenamefont {Seyller}, \citenamefont {Horn},\ and\ \citenamefont {Rotenberg}}]{science1130681}%
  \BibitemOpen
  \bibfield  {author} {\bibinfo {author} {\bibfnamefont {T.}~\bibnamefont {Ohta}}, \bibinfo {author} {\bibfnamefont {A.}~\bibnamefont {Bostwick}}, \bibinfo {author} {\bibfnamefont {T.}~\bibnamefont {Seyller}}, \bibinfo {author} {\bibfnamefont {K.}~\bibnamefont {Horn}}, \ and\ \bibinfo {author} {\bibfnamefont {E.}~\bibnamefont {Rotenberg}},\ }\href {\doibase 10.1126/science.1130681} {\bibfield  {journal} {\bibinfo  {journal} {Science}\ }\textbf {\bibinfo {volume} {313}},\ \bibinfo {pages} {951} (\bibinfo {year} {2006})}\BibitemShut {NoStop}%
\bibitem [{\citenamefont {Blase}\ \emph {et~al.}(1995)\citenamefont {Blase}, \citenamefont {Rubio}, \citenamefont {Louie},\ and\ \citenamefont {Cohen}}]{PhysRevB.51.6868}%
  \BibitemOpen
  \bibfield  {author} {\bibinfo {author} {\bibfnamefont {X.}~\bibnamefont {Blase}}, \bibinfo {author} {\bibfnamefont {A.}~\bibnamefont {Rubio}}, \bibinfo {author} {\bibfnamefont {S.~G.}\ \bibnamefont {Louie}}, \ and\ \bibinfo {author} {\bibfnamefont {M.~L.}\ \bibnamefont {Cohen}},\ }\href {\doibase 10.1103/PhysRevB.51.6868} {\bibfield  {journal} {\bibinfo  {journal} {Phys. Rev. B}\ }\textbf {\bibinfo {volume} {51}},\ \bibinfo {pages} {6868} (\bibinfo {year} {1995})}\BibitemShut {NoStop}%
\bibitem [{\citenamefont {Badrtdinov}\ \emph {et~al.}(2023)\citenamefont {Badrtdinov}, \citenamefont {Rodriguez-Fernandez}, \citenamefont {Grzeszczyk}, \citenamefont {Qiu}, \citenamefont {Vaklinova}, \citenamefont {Huang}, \citenamefont {Hampel}, \citenamefont {Watanabe}, \citenamefont {Taniguchi}, \citenamefont {Jiong}, \citenamefont {Potemski}, \citenamefont {Dreyer}, \citenamefont {Koperski},\ and\ \citenamefont {Rösner}}]{httpsdoi.org10.1002smll.202300144}%
  \BibitemOpen
  \bibfield  {author} {\bibinfo {author} {\bibfnamefont {D.~I.}\ \bibnamefont {Badrtdinov}}, \bibinfo {author} {\bibfnamefont {C.}~\bibnamefont {Rodriguez-Fernandez}}, \bibinfo {author} {\bibfnamefont {M.}~\bibnamefont {Grzeszczyk}}, \bibinfo {author} {\bibfnamefont {Z.}~\bibnamefont {Qiu}}, \bibinfo {author} {\bibfnamefont {K.}~\bibnamefont {Vaklinova}}, \bibinfo {author} {\bibfnamefont {P.}~\bibnamefont {Huang}}, \bibinfo {author} {\bibfnamefont {A.}~\bibnamefont {Hampel}}, \bibinfo {author} {\bibfnamefont {K.}~\bibnamefont {Watanabe}}, \bibinfo {author} {\bibfnamefont {T.}~\bibnamefont {Taniguchi}}, \bibinfo {author} {\bibfnamefont {L.}~\bibnamefont {Jiong}}, \bibinfo {author} {\bibfnamefont {M.}~\bibnamefont {Potemski}}, \bibinfo {author} {\bibfnamefont {C.~E.}\ \bibnamefont {Dreyer}}, \bibinfo {author} {\bibfnamefont {M.}~\bibnamefont {Koperski}}, \ and\ \bibinfo {author} {\bibfnamefont {M.}~\bibnamefont {Rösner}},\ }\href {\doibase https://doi.org/10.1002/smll.202300144} {\bibfield  {journal} {\bibinfo
   {journal} {Small}\ }\textbf {\bibinfo {volume} {19}},\ \bibinfo {pages} {2300144} (\bibinfo {year} {2023})}\BibitemShut {NoStop}%
\bibitem [{\citenamefont {Trolle}\ \emph {et~al.}(2017)\citenamefont {Trolle}, \citenamefont {Pedersen},\ and\ \citenamefont {V{\'e}niard}}]{Trolle2017}%
  \BibitemOpen
  \bibfield  {author} {\bibinfo {author} {\bibfnamefont {M.~L.}\ \bibnamefont {Trolle}}, \bibinfo {author} {\bibfnamefont {T.~G.}\ \bibnamefont {Pedersen}}, \ and\ \bibinfo {author} {\bibfnamefont {V.}~\bibnamefont {V{\'e}niard}},\ }\href {\doibase 10.1038/srep39844} {\bibfield  {journal} {\bibinfo  {journal} {Scientific Reports}\ }\textbf {\bibinfo {volume} {7}},\ \bibinfo {pages} {39844} (\bibinfo {year} {2017})}\BibitemShut {NoStop}%
\bibitem [{\citenamefont {Adeniran}\ and\ \citenamefont {Liu}(2023)}]{PhysRevMaterials.7.054001}%
  \BibitemOpen
  \bibfield  {author} {\bibinfo {author} {\bibfnamefont {O.}~\bibnamefont {Adeniran}}\ and\ \bibinfo {author} {\bibfnamefont {Z.-F.}\ \bibnamefont {Liu}},\ }\href {\doibase 10.1103/PhysRevMaterials.7.054001} {\bibfield  {journal} {\bibinfo  {journal} {Phys. Rev. Mater.}\ }\textbf {\bibinfo {volume} {7}},\ \bibinfo {pages} {054001} (\bibinfo {year} {2023})}\BibitemShut {NoStop}%
\end{thebibliography}

%

\end{document}